\documentclass[aps,prl,preprint,superscriptaddress]{revtex4-2}
\usepackage{amsmath}
\usepackage{graphicx}
\usepackage{bm}%
\usepackage[colorlinks=true,linkcolor=blue]{hyperref}%
\begin{document}
	
	\title{Barrow Holographic Dark Energy with a Ricci Cutoff in Power-Law $f(Q,L_m)$ Gravity: Late-Time Cosmological Dynamics}

    \author{S. Tavakoli}
	\email{samira.Tavakoli@uok.ac.ir}
	
	\author{T. Golanbari}
	\thanks{Corresponding author}
	\email{t.golanbari@gmail.com}
	\email{t.golanbari@uok.ac.ir}
	
	\author{B. Malekolkalami}
	\email{b.malakolkalami@uok.ac.ir}
	
	\author{Kh. Saaidi}
	\email{ksaaidi@uok.ac.ir}

	\affiliation{
		Department of Physics,
		University of Kurdistan,
		P.O. Box 66177-15175,
		Sanandaj,
		Kurdistan,
		Iran
	}

	\date{\today}
	
	\begin{abstract}
		We investigate the late-time cosmological dynamics of Barrow holographic dark energy (BHDE) with the Ricci infrared cutoff in the power-law $f(Q,L_m)$ gravity model $f(Q,L_m)=Q+\alpha Q^n-2L_m$. In this construction, the power-law $Q^n$ contribution modifies the symmetric-teleparallel geometric sector, whereas the Barrow parameter $\Delta$ deforms the holographic energy density through the entropy--area relation. Since the dependence on the matter Lagrangian is linear, the adopted ansatz contains no explicit mixed $Q$--$L_m$ coupling, allowing the effects associated with the modified geometric sector and the Barrow holographic deformation to be identified separately. Combining the modified Friedmann equations with the Ricci-cutoff BHDE density, we derive a closed first-order evolution equation for the normalized Hubble function $E(z)=H(z)/H_0$ and numerically investigate the resulting background cosmology.
		
		The model exhibits a smooth evolution from a matter-dominated regime at positive redshift toward late-time accelerated expansion. The parameters $n$ and $\Delta$ generate distinct signatures in the background observables: increasing $n$ shifts the present dark-energy equation-of-state parameter toward less negative values, increases the acceleration-transition redshift, and decreases the present jerk parameter, whereas increasing $\Delta$ makes both $w_{{\rm DE},0}$ and $q_0$ more negative and increases the transition redshift and the present jerk. Using $q_{0,\rm ref}\simeq-0.527$ to calibrate the fiducial Barrow normalization parameter $\mathcal{C}_{B,\rm ref}$, the resulting background evolution yields $w_{{\rm DE},0}\simeq-1.003$, $z_t\simeq0.442$, and $j_0\simeq1.486$. Two-dimensional maps in the $(n,\Delta)$ plane further reveal that the two deformations act in opposite directions on $w_{{\rm DE},0}$ while shifting the onset of cosmic acceleration in the same direction.
		
		We also compare representative expansion histories with 36 spectroscopic cosmic-chronometer measurements over $0.07\leq z\leq1.965$, incorporating the available DESI covariance information and the asymmetric uncertainty of the measurement at $z=0.8$. This comparison is performed conditionally for fixed benchmark trajectories, without parameter fitting or likelihood optimization, and the resulting conditional Hubble-rate statistic varies only modestly among the representative models. These results provide a consistent background-level realization of Ricci-cutoff Barrow holographic dark energy in power-law $f(Q,L_m)$ gravity and motivate future investigations including cosmological perturbations and joint multi-probe parameter inference.
		
		\vspace{0.5cm}
	\end{abstract}

	\maketitle

	\section{Introduction}\label{sec:introduction}
	
	The discovery of the late-time accelerated expansion of the Universe through observations of Type Ia supernovae fundamentally reshaped the standard cosmological picture \cite{Riess1998Acceleration,Perlmutter1999Acceleration}. Within General Relativity (GR), the simplest explanation is provided by a positive cosmological constant $\Lambda$, and the resulting $\Lambda$CDM scenario offers an economical description of a wide range of cosmological observations, including the cosmic microwave background and the large-scale background evolution \cite{Planck2020Parameters}. Despite this remarkable phenomenological success, the interpretation of $\Lambda$ as vacuum energy gives rise to the long-standing cosmological-constant problem, namely the enormous discrepancy between the observed dark-energy scale and the natural vacuum-energy scales expected from quantum field theory \cite{Weinberg1989CC}. This theoretical difficulty, together with the possibility that the dark sector may possess nontrivial dynamics, continues to motivate alternatives in which cosmic acceleration arises either from additional dynamical degrees of freedom or from a modification of the gravitational sector \cite{Copeland2006DarkEnergy,Clifton2012ModifiedGravity,Aghamohammadi2013ADE}.
	
	Among modified gravitational frameworks, teleparallel and symmetric teleparallel formulations provide geometrically distinct descriptions of gravitation in which torsion or non-metricity, rather than the curvature of the Levi--Civita connection, plays the fundamental dynamical role \cite{Bahamonde2023Teleparallel}. In symmetric teleparallel gravity, both curvature and torsion of the affine connection vanish, while gravity is encoded in the non-metricity tensor. The symmetric teleparallel equivalent of GR, particularly in its coincident formulation, can be expressed in terms of the non-metricity scalar $Q$ \cite{BeltranJimenez2018CGR}. Allowing the gravitational Lagrangian to be a nonlinear function of $Q$ leads to $f(Q)$ gravity, whose cosmological background and perturbative properties have been investigated in a variety of contexts \cite{BeltranJimenez2020fQ}. Observational studies of nonlinear and power-law models further demonstrate that terms proportional to $Q^n$ provide a useful phenomenological setting for testing departures from the standard cosmological background \cite{Mandal2023PowerLawfQ}.
	
	A broader extension is obtained in $f(Q,L_m)$ gravity, where the gravitational function is allowed to depend simultaneously on the non-metricity scalar and the matter Lagrangian $L_m$ \cite{Hazarika2025fQLm}. In its general form, such a dependence permits explicit geometry--matter interactions and can lead to departures from the usual matter-conservation law. The model considered in the present work belongs, however, to a particular and conceptually simpler sector of this theory. We adopt $f(Q,L_m)=Q+\alpha Q^n-2L_m$, for which $f_{QL_m}=0$. Consequently, the model contains no explicit nonminimal $Q$--$L_m$ coupling and instead represents a separable power-law $f(Q)$ sector with linear $L_m$ dependence embedded within the broader $f(Q,L_m)$ framework. This distinction is essential: the modified background dynamics analyzed below originate from the nonlinear $Q^n$ contribution rather than from an explicit exchange between geometry and matter. The exponent $n$ controls the nonlinear structure of the geometric sector, while the amplitude $\alpha$ determines the strength of the corresponding correction. The model therefore provides a simple but sufficiently flexible framework for studying departures from the GR background while retaining a transparent connection with previously investigated power-law $f(Q)$ cosmologies \cite{Mandal2023PowerLawfQ}.
	
	A conceptually different route toward the dark-energy problem arises from the holographic principle. The ultraviolet and infrared sectors of an effective quantum field theory cannot be specified independently when gravitational collapse is taken into account. The UV--IR relation proposed by Cohen, Kaplan, and Nelson imposes an upper bound on the energy contained within a region of characteristic size $L$, thereby providing the theoretical basis for holographic dark-energy models \cite{Cohen1999UVIR}. In the standard holographic dark-energy scenario, the energy density scales as $\rho_{\rm HDE}\propto L^{-2}$, and the cosmological behavior depends crucially on the infrared cutoff chosen to define $L$ \cite{Li2004HDE,Wang2017HDE}. Different choices of cutoff lead to physically different background evolutions and have therefore played a central role in the development of holographic cosmology. Holographic dark energy has also been explored in modified gravitational settings, including scalar--tensor frameworks, where both the cutoff prescription and possible dark-sector interactions influence the resulting cosmological behavior \cite{Saaidi2013HDE}.
	
	An additional generalization of the holographic framework is obtained by modifying the entropy associated with a gravitational horizon. Barrow proposed a deformation of the Bekenstein--Hawking area law motivated by the possibility that quantum-gravitational effects may induce a nontrivial microscopic structure of the horizon \cite{Barrow2020Entropy}. The corresponding entropy is characterized by a deformation parameter $0\leq\Delta\leq1$, with $\Delta=0$ recovering the standard area-law behavior. Applying this generalized entropy to the holographic construction leads to Barrow holographic dark energy (BHDE), whose density scales as $\rho_{\rm DE}=CL^{\Delta-2}$ \cite{Saridakis2020BHDE}. The Barrow parameter thus introduces a deformation of the dark-energy sector that is conceptually independent of the nonlinear gravitational exponent $n$. The cosmological viability of BHDE and the observational sensitivity to $\Delta$ have already been investigated using late-time expansion data \cite{Anagnostopoulos2020BHDE}, providing further motivation for examining its interplay with modified gravitational dynamics.
	
	The infrared cutoff remains a central ingredient of this construction. The original holographic dark-energy model based on the future event horizon involves a nonlocal quantity, whereas alternative prescriptions based on local geometrical combinations can relate the holographic density directly to the instantaneous expansion dynamics \cite{Wang2017HDE}. Of particular relevance is the Ricci prescription proposed by Gao et al., in which the infrared scale is associated with the Ricci curvature of the FLRW metric \cite{Gao2009Ricci}. Related local holographic prescriptions involving combinations of $H^2$ and $\dot H$ were subsequently developed by Granda and Oliveros \cite{GrandaOliveros2008}. In the spatially flat geometry considered here, the Ricci cutoff is defined by $L_R^{-2}=\mathcal{R}/6=2H^2+\dot H$, where $\mathcal{R}$ is the Ricci scalar associated with the Levi--Civita connection of the metric. This qualification is particularly important in symmetric teleparallel gravity: although the affine connection defining the symmetric teleparallel geometry is curvature-free, the metric still possesses a nonvanishing Levi--Civita Ricci scalar. It is this metric curvature scalar, rather than the curvature of the teleparallel affine connection, that enters the infrared prescription. The resulting Barrow density is therefore $\rho_{\rm DE}=C(2H^2+\dot H)^{1-\Delta/2}$, establishing a direct local dependence on both the expansion rate and its first time derivative.
	
	The combination of generalized holographic dark energy with non-metricity-based gravity has attracted increasing attention. Barrow holographic dark energy with a local Granda--Oliveros cutoff has been investigated at the background level, including its expansion history and sensitivity to the Barrow deformation parameter \cite{Oliveros2022BarrowGO}. BHDE has also been studied in $f(Q)$ gravity using the Hubble horizon as the infrared cutoff, including anisotropic cosmologies and solutions obtained from specific time--redshift parametrizations \cite{Koussour2023BHDEfQ}. More recently, a non-interacting BHDE model in $f(Q)$ gravity has been investigated using the Hubble horizon together with a prescribed power-law scale factor \cite{Ajmal2026BHDEfQ}. BHDE has further been examined directly in $f(Q,L_m)$ gravity through a dynamical-system treatment of the linear model $f(Q,L_m)=\alpha Q+\beta L_m$, considering both interacting and non-interacting sectors \cite{Samaddar2026BHDEfQLm}. In addition, a Barrow--Ricci--Gauss--Bonnet holographic construction has recently been explored within $f(Q)$ gravity \cite{Joshi2026BRGBfQ}. A closely related but physically distinct recent study introduced holographic Ricci dark energy in a nonseparable $f(Q,L_m)$ model containing an explicit mixed $QL_m$ term and performed a multi-probe Bayesian analysis \cite{Chokyi2026HRDEfQLm}. These developments demonstrate that Barrow entropy, holographic infrared prescriptions, and symmetric teleparallel gravity can be combined in several inequivalent ways. They also imply that the novelty of a new model must be identified through the precise gravitational sector, the specific infrared cutoff, and the methodology used to determine the cosmological evolution, rather than through the mere coexistence of holographic dark energy and non-metricity.
	
	Against this background, the present work focuses on a distinct construction that combines the separable power-law model $f(Q,L_m)=Q+\alpha Q^n-2L_m$ with Barrow holographic dark energy defined by the pure Ricci cutoff. To the best of our knowledge, this specific combination, together with the self-consistent background evolution developed here, has not previously been analyzed. Rather than prescribing a scale factor, Hubble function, or time--redshift relation, we derive the cosmological evolution directly from the modified Friedmann equations and the Ricci-cutoff BHDE density. Their combination leads to a closed first-order differential equation for the normalized expansion rate $E(z)$, which is solved subject to explicit physical and regularity conditions. Furthermore, the Barrow entropy deformation is used solely to define the holographic dark-energy density, while the gravitational field equations are derived independently from the $f(Q,L_m)$ action. This separation avoids introducing the same entropy deformation simultaneously into both the holographic energy density and a thermodynamic reconstruction of the Friedmann equations.
	
	An important motivation for this construction is that the model contains two physically distinct deformation parameters. The exponent $n$ governs the nonlinear non-metricity correction to the gravitational sector, whereas $\Delta$ controls the deformation of the holographic energy density through the Barrow entropy. Their effects are therefore not interchangeable. We investigate both one-parameter variations and the full joint $(n,\Delta)$ dependence in order to determine how these two modifications influence the late-time cosmological evolution. The resulting dynamics are characterized using the matter, radiation, geometric, and dark-energy density parameters, the dark-energy equation-of-state parameter, the deceleration parameter, and the effective equation of state. We additionally employ the $Om$ diagnostic and the jerk and statefinder parameters \cite{Sahni2003Statefinder,Sahni2008Om}, together with effective energy conditions and the background adiabatic diagnostic $c_a^2=\dot p_{\rm DE}/\dot\rho_{\rm DE}$. The latter is used only as a background quantity and is not interpreted as the physical rest-frame sound speed of dark-energy perturbations. Throughout the analysis, we also impose the physical requirements associated with the Ricci cutoff and the regularity of the modified gravitational equations.
	
	The dynamical analysis is complemented by a direct comparison with observational measurements of the late-time expansion rate. Cosmic chronometers provide estimates of $H(z)$ from the differential aging of passively evolving galaxies and therefore probe the background expansion without requiring the reconstruction of an integrated distance observable \cite{JimenezLoeb2002}. Recent methodological developments include new spectroscopic CC determinations and proof-of-principle photometric extensions of the approach \cite{Borghi2022,Jimenez2023}. To avoid combining physically distinct determinations of the Hubble rate, we employ a curated compilation of 36 spectroscopic cosmic-chronometer measurements spanning $0.07\leq z\leq1.965$, assembled from the original spectroscopic CC analyses rather than from a mixed CC--BAO data set \cite{Simon2005,Stern2010,Zhang2014,Moresco2012,Moresco2015,Moresco2016,Ratsimbazafy2017,Jiao2022,Tomasetti2023,Loubser2025,Loubser2025DESI}. Published covariance information is retained where available, including the correlated DESI measurements, while asymmetric uncertainties are treated without imposing an artificial symmetrization. Possible additional correlations associated with stellar-population modeling are recognized as a limitation of the present background-level comparison \cite{Moresco2020Covariance}.
	
	The observational analysis is performed conditionally for the same representative cosmological trajectories used in the parameter-sensitivity study. For each trajectory, the background parameters are fixed before the theoretical values $H_{\rm th}(z_i)$ are evaluated at the observed redshifts, and the resulting cosmic-chronometer statistic is used only as a conditional discrepancy measure with respect to the common data set. No minimization over $H_0$, $n$, $\Delta$, or the remaining cosmological parameters is carried out, and no best-fit values, likelihood contours, posterior distributions, or parameter constraints are inferred. This construction allows the observational behavior of the representative solutions to be assessed consistently while keeping the scope of the analysis restricted to a controlled background-level comparison.
	
	The remainder of the paper is organized as follows. In Sec.~\ref{sec:fqlm_powerlaw}, we review the field equations of $f(Q,L_m)$ gravity and derive the cosmological equations associated with the power-law model $f(Q,L_m)=Q+\alpha Q^n-2L_m$. Section~\ref{sec:barrow_ricci} introduces Barrow holographic dark energy and specifies the Ricci infrared cutoff. In Sec.~\ref{sec:cosmological_dynamics}, we derive the dimensionless background equations and introduce the cosmological diagnostics used throughout the analysis. Section~\ref{sec:parameter_observational_analysis} studies the one- and two-parameter sensitivity of the background evolution, summarizes the resulting diagnostic behavior, and compares the fixed benchmark trajectories with the spectroscopic cosmic-chronometer sample. Appendix~\ref{app:supplementary_background_diagnostics} presents supplementary background diagnostics for the present-day deceleration parameter and the adiabatic response of the holographic component. Finally, the main conclusions, limitations, and future perspectives are summarized in Sec.~\ref{sec:conclusions}.
	
	\section{$f(Q,L_m)$ Gravity and the Power-Law Model}\label{sec:fqlm_powerlaw}
	
	Symmetric teleparallel gravity describes gravitation through non-metricity while imposing vanishing curvature and torsion on the affine connection. In the coincident formulation, this geometric description provides an alternative representation of general relativity and offers a natural basis for modified theories constructed from nonlinear functions of the non-metricity scalar $Q$ \cite{BeltranJimenez2018CGR,BeltranJimenez2020fQ}. The $f(Q,L_m)$ framework extends this structure by allowing the gravitational Lagrangian to depend on both $Q$ and the matter Lagrangian $L_m$ \cite{Hazarika2025fQLm}. In this section, we introduce the geometric and cosmological relations required for the homogeneous background and then specialize the general theory to the power-law model considered in the present work.
	
	The symmetric teleparallel sector is characterized by vanishing curvature and torsion,
	\begin{equation}\label{eq:stg_constraints}
		R^{\alpha}{}_{\beta\mu\nu}=0,\qquad T^{\alpha}{}_{\mu\nu}=0.
	\end{equation}
	
	The non-metricity tensor is defined as the covariant derivative of the metric,
	\begin{equation}\label{eq:nonmetricity_tensor}
		Q_{\alpha\mu\nu}\equiv\nabla_{\alpha}g_{\mu\nu}.
	\end{equation}
	
	Its two independent traces are
	\begin{equation}\label{eq:nonmetricity_traces}
		Q_{\alpha}\equiv Q_{\alpha\mu}{}^{\mu},\qquad \widetilde{Q}_{\alpha}\equiv Q^{\mu}{}_{\alpha\mu}.
	\end{equation}
	
	With the convention adopted throughout this work, the disformation tensor is given by
	\begin{equation}\label{eq:disformation_tensor}
		L^{\alpha}{}_{\mu\nu}=\frac{1}{2}\left(Q^{\alpha}{}_{\mu\nu}-Q_{\mu}{}^{\alpha}{}_{\nu}-Q_{\nu}{}^{\alpha}{}_{\mu}\right),
	\end{equation}
	
	and the non-metricity conjugate, also referred to as the superpotential, is defined as
	\begin{equation}\label{eq:nonmetricity_conjugate}
		P^{\alpha}{}_{\mu\nu}=-\frac{1}{2}L^{\alpha}{}_{\mu\nu}+\frac{1}{4}\left(Q^{\alpha}-\widetilde{Q}^{\alpha}\right)g_{\mu\nu}-\frac{1}{4}\delta^{\alpha}{}_{(\mu}Q_{\nu)}.
	\end{equation}
	
	The non-metricity scalar is then constructed from the contraction
	\begin{equation}\label{eq:nonmetricity_scalar}
		Q=-Q_{\alpha\mu\nu}P^{\alpha\mu\nu}.
	\end{equation}
	
	We work in units $c=8\pi G=1$ and adopt the action convention used in the original formulation of $f(Q,L_m)$ gravity \cite{Hazarika2025fQLm},
	\begin{equation}\label{eq:fqlm_action}
		S=\int f(Q,L_m)\sqrt{-g}\,d^{4}x.
	\end{equation}
	
	The matter energy-momentum tensor is defined by
	\begin{equation}\label{eq:energy_momentum_tensor}
		T_{\mu\nu}=-\frac{2}{\sqrt{-g}}\frac{\delta\left(\sqrt{-g}L_m\right)}{\delta g^{\mu\nu}}=g_{\mu\nu}L_m-2\frac{\partial L_m}{\partial g^{\mu\nu}}.
	\end{equation}
	
	Variation of the action with respect to the metric gives the metric field equations,
	\begin{widetext}
		\begin{equation}\label{eq:fqlm_field_equations}
			\frac{2}{\sqrt{-g}}\nabla_{\alpha}\left(f_Q\sqrt{-g}P^{\alpha}{}_{\mu\nu}\right)+f_Q\left(P_{\mu\alpha\beta}Q_{\nu}{}^{\alpha\beta}-2Q^{\alpha\beta}{}_{\mu}P_{\alpha\beta\nu}\right)+\frac{1}{2}fg_{\mu\nu}=\frac{1}{2}f_{L_m}\left(g_{\mu\nu}L_m-T_{\mu\nu}\right),
		\end{equation}
	\end{widetext}
	
	where
	\begin{equation}\label{eq:fqlm_derivatives_general}
		f_Q\equiv\frac{\partial f}{\partial Q},\qquad f_{L_m}\equiv\frac{\partial f}{\partial L_m}.
	\end{equation}
	
	The complete metric--affine formulation also contains an equation obtained by varying the independent affine connection. In the present background analysis, we assume that the matter Lagrangian has no explicit dependence on this connection, so the matter hypermomentum vanishes, and we restrict attention to the standard spatially flat FLRW branch in the coincident gauge used in cosmological applications of $f(Q)$ and $f(Q,L_m)$ gravity \cite{BeltranJimenez2020fQ,Hazarika2025fQLm}.
	
	For the cosmological background, we consider a spatially flat Friedmann--Lema\^{i}tre--Robertson--Walker spacetime described by
	\begin{equation}\label{eq:flrw_metric}
		ds^{2}=-dt^{2}+a^{2}(t)\left(dx^{2}+dy^{2}+dz^{2}\right),
	\end{equation}
	
	where $a(t)$ denotes the scale factor and $H\equiv\dot a/a$ is the Hubble parameter. The cosmic medium is described by a perfect fluid,
	\begin{equation}\label{eq:perfect_fluid_tensor}
		T_{\mu\nu}=(\rho+p)u_{\mu}u_{\nu}+pg_{\mu\nu},
	\end{equation}
	
	with $u^{\mu}u_{\mu}=-1$. Following the cosmological formulation of $f(Q,L_m)$ gravity, we adopt the matter Lagrangian
	\begin{equation}\label{eq:matter_lagrangian_choice}
		L_m=-\rho.
	\end{equation}
	
	This choice fixes the matter-fluid convention used throughout the analysis. For the particular model introduced below, however, the dependence on $L_m$ is linear and $f_{L_m}$ is constant. As a result, the explicit dependence on the chosen representation of $L_m$ cancels from the homogeneous background equations. On the standard spatially flat coincident-gauge FLRW branch specified above, the non-metricity scalar becomes \cite{BeltranJimenez2020fQ,Hazarika2025fQLm}
	\begin{equation}\label{eq:Q_FLRW}
		Q=6H^{2}.
	\end{equation}
	
	The two generalized Friedmann equations of $f(Q,L_m)$ gravity can then be written as \cite{Hazarika2025fQLm}
	\begin{equation}\label{eq:general_friedmann_1}
		3H^{2}=\frac{1}{4f_Q}\left[f-f_{L_m}(\rho+L_m)\right],
	\end{equation}
	
	and
	\begin{equation}\label{eq:general_friedmann_2}
		\dot H+3H^{2}+H\frac{\dot f_Q}{f_Q}=\frac{1}{4f_Q}\left[f+f_{L_m}(p-L_m)\right].
	\end{equation}
	
	Subtracting Eq.~(\ref{eq:general_friedmann_1}) from Eq.~(\ref{eq:general_friedmann_2}) gives the useful background relation
	\begin{equation}\label{eq:general_fqH_relation}
		\frac{d}{dt}\left(f_QH\right)=\frac{f_{L_m}}{4}(\rho+p).
	\end{equation}
	
	We now specialize the general framework to the power-law model
	\begin{equation}\label{eq:powerlaw_model}
		f(Q,L_m)=Q+\alpha Q^{n}-2L_m,
	\end{equation}
	
	where $\alpha$ determines the amplitude of the nonlinear geometric correction and $n$ controls its power-law dependence on $Q$. The derivatives entering the background field equations are
	\begin{equation}\label{eq:powerlaw_derivatives}
		f_Q=1+\alpha nQ^{n-1},\qquad f_{QQ}=\alpha n(n-1)Q^{n-2},\qquad f_{L_m}=-2,\qquad f_{QL_m}=0.
	\end{equation}
	
	The last relation in Eq.~(\ref{eq:powerlaw_derivatives}) shows that the adopted ansatz contains no explicit mixed $Q$--$L_m$ coupling. Defining $F(Q)\equiv Q+\alpha Q^n$, the action is separable as $S=\int\sqrt{-g}\,[F(Q)-2L_m]\,d^4x$. Up to an overall nonzero normalization of the complete action, which does not affect the Euler--Lagrange equations, this has the same dynamical structure as the conventional minimally coupled $f(Q)$ action $\int\sqrt{-g}\,[-F(Q)/2+L_m]\,d^4x$ \cite{BeltranJimenez2020fQ}. Thus, the present model represents the minimally coupled power-law $f(Q)$ sector embedded in the broader $f(Q,L_m)$ framework: its modified background dynamics arise from the nonlinear term $\alpha Q^n$, not from a nonminimal matter--geometry interaction.
	
	Substituting Eq.~(\ref{eq:powerlaw_model}) into Eq.~(\ref{eq:general_friedmann_1}) and using Eq.~(\ref{eq:Q_FLRW}) yields
	\begin{equation}\label{eq:powerlaw_friedmann_1}
		3H^{2}=\rho-\frac{\alpha}{2}(2n-1)Q^{n}=\rho-\frac{\alpha}{2}(2n-1)(6H^{2})^{n}.
	\end{equation}
	
	For later convenience, we introduce the auxiliary geometric correction
	\begin{equation}\label{eq:rhoQ_definition}
		\rho_Q\equiv\frac{\alpha}{2}(2n-1)Q^{n}.
	\end{equation}
	
	The first modified Friedmann equation can therefore be expressed in the compact form
	\begin{equation}\label{eq:powerlaw_friedmann_compact}
		3H^{2}=\rho-\rho_Q.
	\end{equation}
	
	The quantity $\rho_Q$ is introduced only as a convenient representation of the nonlinear geometric contribution. It should not be interpreted as an independently conserved material fluid, since the actual contribution of this term to the right-hand side of Eq.~(\ref{eq:powerlaw_friedmann_compact}) appears with the opposite sign.
	
	The evolution equation for the Hubble parameter follows directly from Eq.~(\ref{eq:general_fqH_relation}). Using
	\begin{equation}\label{eq:powerlaw_fqdot}
		\dot f_Q=\alpha n(n-1)Q^{n-2}\dot Q,\qquad \dot Q=12H\dot H,
	\end{equation}
	
	we obtain
	\begin{equation}\label{eq:powerlaw_hdot_rhop}
		\dot H=-\frac{\rho+p}{2\left[1+\alpha n(2n-1)Q^{n-1}\right]}.
	\end{equation}
	
	Combining Eq.~(\ref{eq:powerlaw_hdot_rhop}) with Eq.~(\ref{eq:powerlaw_friedmann_1}), the second background equation can equivalently be written in Raychaudhuri form as
	\begin{equation}\label{eq:powerlaw_raychaudhuri}
		2\dot H+3H^{2}=-p-2\alpha n(2n-1)Q^{n-1}\dot H+\frac{\alpha}{2}(1-2n)Q^{n}.
	\end{equation}
	
	Equations~(\ref{eq:powerlaw_friedmann_1}) and (\ref{eq:powerlaw_raychaudhuri}) constitute the two independent background equations of the power-law model and will later be combined with the Barrow holographic dark-energy sector.
	
	For the present model, the dependence on $L_m$ is linear and $f_{L_m}=-2$ is constant. The standard background conservation equation is therefore recovered consistently. Differentiating Eq.~(\ref{eq:powerlaw_friedmann_compact}) and using Eqs.~(\ref{eq:powerlaw_hdot_rhop}) and (\ref{eq:powerlaw_fqdot}) gives
	\begin{equation}\label{eq:powerlaw_total_continuity}
		\dot\rho+3H(\rho+p)=0.
	\end{equation}
	
	We decompose the total cosmic density and pressure as
	\begin{equation}\label{eq:cosmic_component_decomposition}
		\rho=\rho_m+\rho_r+\rho_{DE},\qquad p=p_m+p_r+p_{DE},\qquad p_m=0,\qquad p_r=\frac{1}{3}\rho_r,\qquad p_{DE}=w_{DE}\rho_{DE}.
	\end{equation}
	
	If matter, radiation, and dark energy are assumed to be mutually noninteracting, Eq.~(\ref{eq:powerlaw_total_continuity}) separates into
	\begin{equation}\label{eq:matter_continuity}
		\dot\rho_m+3H\rho_m=0,
	\end{equation}
	
	\begin{equation}\label{eq:radiation_continuity}
		\dot\rho_r+4H\rho_r=0,
	\end{equation}
	
	and
	\begin{equation}\label{eq:darkenergy_continuity}
		\dot\rho_{DE}+3H(1+w_{DE})\rho_{DE}=0,
	\end{equation}
	
	where $w_{DE}\equiv p_{DE}/\rho_{DE}$. The explicit form of $\rho_{DE}$ will be supplied by Barrow holographic dark energy in the next section.
	
	We define the standard dimensionless density parameters by
	\begin{equation}\label{eq:density_parameters_standard}
		\Omega_m\equiv\frac{\rho_m}{3H^{2}},\qquad \Omega_r\equiv\frac{\rho_r}{3H^{2}},\qquad \Omega_{DE}\equiv\frac{\rho_{DE}}{3H^{2}},
	\end{equation}
	
	and introduce the corresponding dimensionless geometric correction
	\begin{equation}\label{eq:OmegaQ_definition}
		\Omega_Q\equiv\frac{\rho_Q}{3H^{2}}=\alpha(2n-1)Q^{n-1}.
	\end{equation}
	
	The first Friedmann equation then leads to the closure relation
	\begin{equation}\label{eq:powerlaw_closure_relation}
		1=\Omega_m+\Omega_r+\Omega_{DE}-\Omega_Q.
	\end{equation}
	
	Using Eqs.~(\ref{eq:powerlaw_hdot_rhop}) and (\ref{eq:OmegaQ_definition}), the normalized Hubble derivative becomes
	\begin{equation}\label{eq:powerlaw_hdot_dimensionless}
		\frac{\dot H}{H^{2}}=-\frac{3\Omega_m+4\Omega_r+3(1+w_{DE})\Omega_{DE}}{2(1+n\Omega_Q)}.
	\end{equation}
	
	Equation~(\ref{eq:powerlaw_hdot_dimensionless}) will play a central role in the cosmological diagnostics derived in the following sections. A regular Hubble evolution requires
	\begin{equation}\label{eq:regularity_condition_hdot}
		1+n\Omega_Q\neq0,
	\end{equation}
	
	since the vanishing of this denominator would make Eq.~(\ref{eq:powerlaw_hdot_dimensionless}) singular. In addition, the generalized Friedmann equations contain the factor $1/f_Q$, and therefore a regular background solution must satisfy
	\begin{equation}\label{eq:regularity_condition_fq}
		f_Q=1+\alpha nQ^{n-1}\neq0.
	\end{equation}
	
	For $n\neq1/2$, Eq.~(\ref{eq:OmegaQ_definition}) allows the latter condition to be expressed as
	\begin{equation}\label{eq:regularity_condition_fq_omega}
		1+\frac{n}{2n-1}\Omega_Q\neq0.
	\end{equation}
	
	Because $Q$ has dimensions of $H^{2}$, the coefficient $\alpha$ has dimensions
	\[
	[\alpha]=[Q]^{1-n}=[H]^{2(1-n)},
	\]
	and is therefore dimensionful for $n\neq1$. To separate this dimensional normalization from the power-law index, we introduce the present-day non-metricity scale and the dimensionless Hubble function,
	\begin{equation}\label{eq:Q0_E_definition}
		Q_0\equiv6H_0^{2},\qquad E(z)\equiv\frac{H(z)}{H_0},
	\end{equation}
	together with the dimensionless coefficient
	\begin{equation}\label{eq:alpha_bar_definition}
		\bar\alpha\equiv\alpha Q_0^{n-1}.
	\end{equation}
	
	The present value of the geometric correction is then
	\begin{equation}\label{eq:OmegaQ0_definition}
		\Omega_{Q0}=(2n-1)\bar\alpha.
	\end{equation}
	
	For fixed nonzero $\Omega_{Q0}$ and $n\neq1/2$, this relation gives $\bar\alpha=\Omega_{Q0}/(2n-1)$. At $n=1/2$, no finite $\bar\alpha$ can represent a fixed nonzero $\Omega_{Q0}$ because the factor $2n-1$ vanishes. This is a degeneracy of the fixed-$\Omega_{Q0}$ parametrization, not a singularity of the homogeneous cosmological background.
	
	Using $Q/Q_0=E^2(z)$, its redshift evolution becomes
	\begin{equation}\label{eq:OmegaQ_E_relation}
		\Omega_Q(z)=\Omega_{Q0}E^{2(n-1)}(z),
	\end{equation}
	
	while Eq.~(\ref{eq:rhoQ_definition}) can equivalently be written as
	\begin{equation}\label{eq:rhoQ_E_relation}
		\frac{\rho_Q(z)}{3H_0^{2}}=\Omega_{Q0}E^{2n}(z).
	\end{equation}
	
	These relations provide a dimensionless representation of the nonlinear geometric correction and will be useful in the numerical treatment of the cosmological equations.
	
	Three particular values of $n$ provide useful reference limits. For $n=0$, the nonlinear term becomes a constant and the model reduces to $f(Q,L_m)=Q+\alpha-2L_m$. In this case, Eq.~(\ref{eq:powerlaw_friedmann_1}) gives $3H^2=\rho+\alpha/2$, so the additional contribution behaves at the homogeneous level as an effective cosmological constant with $\Lambda_{\rm eff}=\alpha/2$. For $n=1/2$, the factor $2n-1$ vanishes, implying $\rho_Q=0$ and removing the nonlinear correction from the homogeneous Friedmann and Raychaudhuri equations. This branch is therefore degenerate with the standard homogeneous background at the level considered here, even though the underlying function $f_Q$ can remain different from unity. As emphasized above, $n=1/2$ is not a cosmological singularity; it is excluded only from scans that hold a nonzero $\Omega_{Q0}$ fixed, because that parametrization would require a divergent $\bar\alpha$. For $n=1$, the model becomes $f(Q,L_m)=(1+\alpha)Q-2L_m$, and the modification reduces to a constant rescaling of the gravitational sector. Consequently, genuinely nontrivial power-law corrections to the homogeneous expansion arise for values of $n$ away from these special cases.
	
	Finally, the standard background equations are continuously recovered when the nonlinear correction is removed. For fixed $n\neq1/2$, the limit $\alpha\rightarrow0$ is equivalent to $\Omega_{Q0}\rightarrow0$, and Eqs.~(\ref{eq:powerlaw_friedmann_1}), (\ref{eq:powerlaw_raychaudhuri}), and (\ref{eq:powerlaw_total_continuity}) reduce to
	\begin{equation}\label{eq:GR_background_limit}
		3H^{2}=\rho,\qquad 2\dot H+3H^{2}=-p,\qquad \dot\rho+3H(\rho+p)=0.
	\end{equation}
	
	This limit provides a direct consistency check for both the analytical derivations and the subsequent numerical implementation. In the action convention of Eq.~(\ref{eq:fqlm_action}), the $\alpha=0$ theory, $f(Q,L_m)=Q-2L_m$, differs from the conventional STEGR normalization only by an overall nonzero factor and therefore yields the same background field equations. In the next section, Barrow holographic dark energy with the Ricci infrared cutoff will be introduced and then combined with the background equations derived above.
	
	\section{Barrow Holographic Dark Energy with Ricci Cutoff}\label{sec:barrow_ricci}
	
	The holographic description of dark energy is based on the connection between ultraviolet and infrared scales in an effective quantum field theory. Requiring that the total energy contained in a region of characteristic size $L$ does not exceed the mass of a black hole of the same size leads to an infrared-dependent vacuum-energy scale \cite{Cohen1999UVIR}. When this idea is applied to cosmology, the corresponding dark-energy density takes the standard holographic form $\rho_{DE}\propto L^{-2}$, where the choice of the infrared cutoff $L$ determines the cosmological dynamics \cite{Li2004HDE}. Generalized horizon entropies modify this scaling and therefore provide a natural way to construct extended holographic dark-energy models.
	
	Barrow proposed a deformation of the usual black-hole entropy motivated by the possibility that quantum-gravitational effects generate a highly intricate structure on the horizon \cite{Barrow2020Entropy}. The resulting entropy is written as
	\begin{equation}\label{eq:barrow_entropy}
		S_{B}=\left(\frac{A}{A_{0}}\right)^{1+\Delta/2},
	\end{equation}
	
	where $A$ denotes the horizon area, $A_{0}$ is the Planck area, and $\Delta$ is the Barrow deformation parameter. In the original construction, the parameter lies in the interval
	\begin{equation}\label{eq:barrow_delta_range}
		0\leq\Delta\leq1,
	\end{equation}
	
	with $\Delta=0$ recovering the standard Bekenstein--Hawking area law, while $\Delta=1$ corresponds to the maximal deformation considered in the Barrow framework \cite{Barrow2020Entropy,Saridakis2020BHDE}. Thus, $\Delta$ directly measures the departure of the horizon entropy from the standard area scaling.
	
	Replacing the Bekenstein--Hawking entropy in the usual holographic construction by the Barrow entropy leads to Barrow holographic dark energy \cite{Saridakis2020BHDE}. The corresponding energy density can be expressed as
	\begin{equation}\label{eq:bhde_general_density}
		\rho_{DE}=C\,L^{\Delta-2},
	\end{equation}
	
	where $C$ is a positive normalization constant whose dimensions are fixed by the requirement that $\rho_{DE}$ has the dimensions of an energy density. In natural units with $\hbar=c=1$ while retaining the Planck scale explicitly, an energy density has mass dimension four, and Eq.~(\ref{eq:bhde_general_density}) therefore implies $[C]=[L]^{-2-\Delta}$. In the units $8\pi G=1$ used for the cosmological background equations, the gravitational density scale is absorbed so that $\rho$ has the same dimensions as $H^2$; correspondingly, the normalization entering the background equations has effective dimension $[C]=[H]^{\Delta}$. This convention makes the dimensionless normalization introduced in Sec.~\ref{sec:cosmological_dynamics} transparent. Equation~(\ref{eq:bhde_general_density}) reduces continuously to the standard holographic relation in the undeformed limit,
	\begin{equation}\label{eq:bhde_standard_limit}
		\left.\rho_{DE}\right|_{\Delta=0}=C\,L^{-2}.
	\end{equation}
	
	At $\Delta=0$, the conventional normalization $C=3c^{2}M_{\rm p}^{2}$ reproduces the usual holographic dark-energy density,
	\begin{equation}\label{eq:standard_hde_density}
		\rho_{HDE}=3c^{2}M_{\rm p}^{2}L^{-2},
	\end{equation}
	
	where $M_{\rm p}$ is the reduced Planck mass and $c$ is the standard dimensionless holographic parameter \cite{Li2004HDE,Saridakis2020BHDE}. In the units $8\pi G=1$ adopted in Sec.~\ref{sec:fqlm_powerlaw}, $M_{\rm p}=1$, while the Barrow-sector normalization is retained explicitly through $C$. In Sec.~\ref{sec:cosmological_dynamics}, this quantity is recast as the dimensionless combination $\mathcal{C}_{B}=C/(3H_0^{\Delta})$.
	
	A central ingredient of any holographic dark-energy model is the specification of the infrared cutoff. In the present work, we choose a Ricci-type cutoff rather than the future event horizon. This choice is local in the background variables and is determined by the instantaneous expansion rate and its first time derivative. The Ricci dark-energy construction was originally motivated by identifying the infrared scale with the characteristic curvature scale of the spatially flat FLRW spacetime \cite{Gao2009Ricci}. With the curvature-sign convention adopted here, the Ricci scalar associated with the Levi--Civita connection of the metric in Eq.~(\ref{eq:flrw_metric}) is
	\begin{equation}\label{eq:levicivita_ricci_scalar}
		\mathcal{R}=6\left(2H^{2}+\dot H\right).
	\end{equation}
	
	The symbol $\mathcal{R}$ in Eq.~(\ref{eq:levicivita_ricci_scalar}) denotes the Ricci scalar of the Levi-Civita connection constructed from the metric. This should not be confused with the curvature of the independent affine connection used in symmetric teleparallel gravity, which vanishes by construction according to Eq.~(\ref{eq:stg_constraints}). The Ricci cutoff therefore introduces no inconsistency with the symmetric teleparallel geometric conditions; it is used only as a local metric-based infrared scale for the holographic dark-energy sector.
	
	Following the usual Ricci holographic prescription, we define the infrared cutoff by
	\begin{equation}\label{eq:ricci_cutoff_definition}
		L_{R}^{-2}\equiv\frac{\mathcal{R}}{6}=2H^{2}+\dot H.
	\end{equation}
	
	The numerical factor $1/6$ in Eq.~(\ref{eq:ricci_cutoff_definition}) is a convention and could equivalently be absorbed into the normalization constant of the holographic energy density. The present definition is particularly convenient because it reproduces the standard Ricci dark-energy combination $2H^{2}+\dot H$ directly \cite{Gao2009Ricci}.
	
	Substituting the Ricci cutoff of Eq.~(\ref{eq:ricci_cutoff_definition}) into the Barrow holographic density in Eq.~(\ref{eq:bhde_general_density}) gives
	\begin{equation}\label{eq:bhde_ricci_density}
		\rho_{DE}=C\left(2H^{2}+\dot H\right)^{1-\Delta/2}.
	\end{equation}
	
	Equation~(\ref{eq:bhde_ricci_density}) is the fundamental dark-energy relation adopted in the remainder of this work. It contains two distinct ingredients: the exponent $\Delta$, which measures the deformation of the horizon entropy, and the curvature combination $2H^{2}+\dot H$, which specifies the Ricci infrared scale. The normalization constant $C$ controls the overall amplitude of the holographic component and will later be expressed in a dimensionless form suitable for the background numerical analysis.
	
	Using the definition of the deceleration parameter,
	\begin{equation}\label{eq:deceleration_definition_preliminary}
		q\equiv-1-\frac{\dot H}{H^{2}},
	\end{equation}
	
	the Ricci combination can also be written as
	\begin{equation}\label{eq:ricci_combination_q}
		2H^{2}+\dot H=H^{2}(1-q).
	\end{equation}
	
	On the expanding cosmological branch considered throughout this work, $H>0$, so that $(H^2)^{1-\Delta/2}=H^{2-\Delta}$. Consequently, the Barrow holographic density may equivalently be expressed as
	\begin{equation}\label{eq:bhde_density_q_form}
		\rho_{DE}=C\,H^{2-\Delta}(1-q)^{1-\Delta/2}.
	\end{equation}
	
	Equation~(\ref{eq:bhde_density_q_form}) is useful for interpreting the connection between the holographic energy density and the kinematics of the cosmic expansion, although the numerical evolution will be formulated directly in terms of $H$ and its derivative.
	
	For a generic noninteger value of the exponent $1-\Delta/2$, a real-valued Barrow holographic density is ensured when the base of the power in Eq.~(\ref{eq:bhde_ricci_density}) is non-negative. The limiting case $2H^2+\dot H=0$ corresponds to $L_R^{-2}=0$, and hence to an infinite Ricci infrared length. In the present analysis we restrict attention to the positive, finite Ricci-cutoff branch used in the subsequent numerical inversion. The physical background solutions are therefore required to satisfy
	\begin{equation}\label{eq:ricci_cutoff_reality_condition}
		2H^{2}+\dot H>0,
	\end{equation}
	
	or equivalently,
	\begin{equation}\label{eq:ricci_cutoff_reality_q}
		q<1.
	\end{equation}
	
	This condition will be imposed as a regularity requirement in the subsequent numerical analysis. It is naturally satisfied during the standard matter-dominated and accelerating epochs and ensures a real, positive, and finite Ricci infrared scale on the branch considered here.
	
	The undeformed limit provides an important consistency check. Taking $\Delta\rightarrow0$ in Eq.~(\ref{eq:bhde_ricci_density}) gives
	\begin{equation}\label{eq:bhde_ricci_delta_zero}
		\left.\rho_{DE}\right|_{\Delta=0}=C\left(2H^{2}+\dot H\right).
	\end{equation}
	
	With $C=3c^{2}M_{\rm p}^{2}$, Eq.~(\ref{eq:bhde_ricci_delta_zero}) reduces to the standard Ricci holographic dark-energy form,
	\begin{equation}\label{eq:standard_ricci_hde_limit}
		\rho_{RDE}=3c^{2}M_{\rm p}^{2}\left(2H^{2}+\dot H\right),
	\end{equation}
	
	up to the conventional notation adopted for the dimensionless Ricci holographic coefficient \cite{Gao2009Ricci}. Hence, the model possesses a continuous limit in which the Barrow deformation is removed while the Ricci infrared prescription is retained.
	
	It is useful to emphasize that the role of Barrow entropy in the present construction is restricted to the dark-energy sector. The modified gravitational equations remain those derived from the power-law $f(Q,L_m)$ model in Sec.~\ref{sec:fqlm_powerlaw}; we do not independently modify the Friedmann equations through a thermodynamic derivation based on Barrow entropy. This separation avoids counting the same entropy modification simultaneously as a correction to the gravitational field equations and as a holographic energy component.
	
	The relations derived in this section complete the definition of the dark-energy sector. In the next section, Eq.~(\ref{eq:bhde_ricci_density}) will be combined with the modified Friedmann and Hubble-evolution equations obtained in Sec.~\ref{sec:fqlm_powerlaw}. This will provide the closed background system from which the dark-energy density parameter, equation-of-state parameter, deceleration parameter, and the remaining cosmological diagnostics will be obtained.
	
	\section{Cosmological Dynamics and Diagnostics}\label{sec:cosmological_dynamics}
	
	The gravitational sector developed in Sec.~\ref{sec:fqlm_powerlaw} and the Barrow holographic dark-energy density introduced in Sec.~\ref{sec:barrow_ricci} can now be combined to obtain a closed system for the homogeneous cosmological evolution. Throughout this section, matter, radiation, and Barrow holographic dark energy are assumed to be mutually noninteracting, so that the conservation equations derived in Sec.~\ref{sec:fqlm_powerlaw} remain valid. The analytical evolution equations are written in terms of the redshift $z$, defined through $1+z=a_0/a$ with $a_0=1$, on the branch $z>-1$. In the numerical implementation, the equivalent e-fold variable $x=\ln a=-\ln(1+z)$ is used for the integration, while the background observables are reconstructed as functions of $z$.
	
	\subsection{Background Cosmological Evolution}\label{subsec:background_evolution}
	
	Using $d/dt=-(1+z)H\,d/dz$ together with $E(z)=H(z)/H_0$, the time derivative of the Hubble parameter can be written as
	\begin{equation}\label{eq:Hdot_redshift}
		\dot H=-H_0^2(1+z)E E_{,z},
	\end{equation}
	
	where $E_{,z}\equiv dE/dz$. The Ricci combination appearing in the Barrow holographic density therefore becomes
	\begin{equation}\label{eq:ricci_combination_dimensionless}
		2H^2+\dot H=H_0^2\mathcal{B}(z),
	\end{equation}
	
	where we have introduced
	\begin{equation}\label{eq:B_dimensionless_definition}
		\mathcal{B}(z)\equiv2E^2-(1+z)E E_{,z}.
	\end{equation}
	
	For compactness, we define
	\begin{equation}\label{eq:barrow_exponent_nu}
		\nu\equiv1-\frac{\Delta}{2}=\frac{2-\Delta}{2},
	\end{equation}
	
	which satisfies $1/2\leq\nu\leq1$ for $0\leq\Delta\leq1$. In the $8\pi G=1$ convention adopted here, the normalization constant $C$ in Eq.~(\ref{eq:bhde_ricci_density}) has effective dimension $[C]=[H]^\Delta$ and is therefore dimensionful for $\Delta\neq0$. We therefore introduce the dimensionless parameter
	\begin{equation}\label{eq:Cbarrow_dimensionless}
		\mathcal{C}_{B}\equiv\frac{C}{3H_0^\Delta}.
	\end{equation}
	
	The Barrow holographic dark-energy density can consequently be written as
	\begin{equation}\label{eq:rhoDE_dimensionless}
		\frac{\rho_{DE}}{3H_0^2}=\mathcal{C}_{B}\mathcal{B}^{\nu}.
	\end{equation}
	
	Since matter and radiation obey their standard conservation equations, their densities evolve according to
	\begin{equation}\label{eq:matter_radiation_redshift}
		\frac{\rho_m}{3H_0^2}=\Omega_{m0}(1+z)^3,\qquad \frac{\rho_r}{3H_0^2}=\Omega_{r0}(1+z)^4.
	\end{equation}
	
	Using Eq.~(\ref{eq:rhoQ_E_relation}) for the nonlinear geometric correction, the modified Friedmann equation becomes
	\begin{equation}\label{eq:dimensionless_friedmann_bhde}
		E^2=\Omega_{m0}(1+z)^3+\Omega_{r0}(1+z)^4+\mathcal{C}_{B}\mathcal{B}^{\nu}-\Omega_{Q0}E^{2n}.
	\end{equation}
	
	For convenience, we define
	\begin{equation}\label{eq:D_background_definition}
		\mathcal{D}(z,E)\equiv E^2-\Omega_{m0}(1+z)^3-\Omega_{r0}(1+z)^4+\Omega_{Q0}E^{2n}.
	\end{equation}
	
	Equation~(\ref{eq:dimensionless_friedmann_bhde}) then gives
	\begin{equation}\label{eq:B_D_relation}
		\mathcal{C}_{B}\mathcal{B}^{\nu}=\mathcal{D}.
	\end{equation}
	
	For $\mathcal{C}_{B}>0$ and $\mathcal{D}>0$, the real branch is
	\begin{equation}\label{eq:B_explicit_solution}
		\mathcal{B}=\left(\frac{\mathcal{D}}{\mathcal{C}_{B}}\right)^{2/(2-\Delta)}.
	\end{equation}
	
	Combining Eqs.~(\ref{eq:B_dimensionless_definition}) and (\ref{eq:B_explicit_solution}), the background dynamics reduces to the first-order differential equation
	\begin{equation}\label{eq:E_redshift_ODE}
		E_{,z}=\frac{2E^2-\left(\mathcal{D}/\mathcal{C}_{B}\right)^{2/(2-\Delta)}}{(1+z)E}.
	\end{equation}
	
	The normalization of the Hubble function is fixed by $E(0)=1$.
	Equation~(\ref{eq:E_redshift_ODE}) is the central evolution equation of the background analysis. In particular, the Ricci-cutoff Barrow density is incorporated directly into a first-order equation for $E(z)$, so that no independent differential equation for the dark-energy equation-of-state parameter is required.
	
	At the present epoch, the closure relation in Eq.~(\ref{eq:powerlaw_closure_relation}) gives
	\begin{equation}\label{eq:OmegaDE0_closure}
		\Omega_{DE0}=1-\Omega_{m0}-\Omega_{r0}+\Omega_{Q0}.
	\end{equation}
	
	Since $\mathcal{D}(0,1)=\Omega_{DE0}$, the present deceleration parameter satisfies
	\begin{equation}\label{eq:q0_Cbarrow_relation}
		q_0=1-\left(\frac{\Omega_{DE0}}{\mathcal{C}_{B}}\right)^{2/(2-\Delta)}.
	\end{equation}
	
	Equivalently,
	\begin{equation}\label{eq:Cbarrow_q0_relation}
		\mathcal{C}_{B}=\frac{\Omega_{DE0}}{(1-q_0)^{1-\Delta/2}}.
	\end{equation}
	
	Hence, $\mathcal{C}_{B}$ and $q_0$ are not independent once $\Omega_{m0}$, $\Omega_{r0}$, $\Omega_{Q0}$, and $\Delta$ are specified. Either $\mathcal{C}_{B}$ or $q_0$ can therefore be adopted as a background parameter, but they should not be varied independently.
	
	A convenient parameter set for the background evolution is
	\begin{equation}\label{eq:background_parameter_set}
		\boldsymbol{\Theta}_{\rm bg}=\{\Omega_{m0},\Omega_{r0},\Omega_{Q0},\mathcal{C}_{B},n,\Delta\}.
	\end{equation}
	
	The Hubble constant $H_0$ fixes the dimensional normalization through $H(z)=H_0E(z)$. In the parameter-sensitivity analysis performed in this work, when the $(n,\Delta)$ plane is explored, all remaining background quantities in Eq.~(\ref{eq:background_parameter_set}) are held fixed at explicitly stated benchmark values; no statistical marginalization or profiling over additional background parameters is carried out.
	
	The numerical solutions are restricted to the real and regular branch of the model. In addition to the conditions obtained in Sec.~\ref{sec:fqlm_powerlaw}, we require
	\begin{equation}\label{eq:background_physical_conditions}
		E>0,\qquad \mathcal{C}_{B}>0,\qquad \mathcal{D}>0,\qquad \mathcal{B}>0,\qquad f_Q\neq0,\qquad 1+n\Omega_Q\neq0.
	\end{equation}
	
	The requirements $\mathcal{D}>0$ and $\mathcal{B}>0$ select the positive Barrow-density branch and a real, positive, and finite Ricci infrared scale throughout the redshift interval considered. On an exact solution these quantities are algebraically linked through Eq.~(\ref{eq:B_D_relation}), but both are monitored in the numerical implementation as separate numerical consistency checks.
	
	\subsection{Density Parameters and Dark-Energy Equation of State}\label{subsec:density_eos}
	
	Once Eq.~(\ref{eq:E_redshift_ODE}) has been solved, the matter and radiation density parameters follow as
	\begin{equation}\label{eq:Omega_m_redshift}
		\Omega_m(z)=\frac{\Omega_{m0}(1+z)^3}{E^2(z)},
	\end{equation}
	
	and
	\begin{equation}\label{eq:Omega_r_redshift}
		\Omega_r(z)=\frac{\Omega_{r0}(1+z)^4}{E^2(z)}.
	\end{equation}
	
	The nonlinear geometric correction evolves according to
	\begin{equation}\label{eq:Omega_Q_redshift_section4}
		\Omega_Q(z)=\Omega_{Q0}E^{2(n-1)}(z).
	\end{equation}
	
	The Barrow holographic dark-energy density parameter follows directly from the closure relation,
	\begin{equation}\label{eq:Omega_DE_closure}
		\Omega_{DE}(z)=1-\Omega_m(z)-\Omega_r(z)+\Omega_Q(z).
	\end{equation}
	
	An equivalent expression follows directly from the holographic density,
	\begin{equation}\label{eq:Omega_DE_B_form}
		\Omega_{DE}(z)=\frac{\mathcal{C}_{B}\mathcal{B}^{\nu}}{E^2}.
	\end{equation}
	
	Since $\mathcal{B}=E^2(1-q)$, Eq.~(\ref{eq:Omega_DE_B_form}) can also be written as
	\begin{equation}\label{eq:Omega_DE_q_form}
		\Omega_{DE}(z)=\mathcal{C}_{B}E^{-\Delta}(1-q)^{1-\Delta/2}.
	\end{equation}
	
	Equations~(\ref{eq:Omega_DE_closure}) and (\ref{eq:Omega_DE_B_form}) provide an internal consistency test for the numerical integration, since they must yield the same $\Omega_{DE}(z)$ along a given solution.
	
	The equation-of-state parameter of the Barrow holographic component is defined as
	\begin{equation}\label{eq:wDE_definition}
		w_{DE}\equiv\frac{p_{DE}}{\rho_{DE}}.
	\end{equation}
	
	Since the dark-energy component is assumed to be separately conserved, Eq.~(\ref{eq:darkenergy_continuity}) gives
	\begin{equation}\label{eq:wDE_continuity_form}
		w_{DE}=-1+\frac{1+z}{3}\frac{d\ln\rho_{DE}}{dz}.
	\end{equation}
	
	Direct differentiation of Eq.~(\ref{eq:bhde_ricci_density}) would introduce a second derivative of the Hubble function. A more convenient relation follows from the modified Hubble-evolution equation in Eq.~(\ref{eq:powerlaw_hdot_dimensionless}). Using Eq.~(\ref{eq:Hdot_redshift}), one obtains
	\begin{equation}\label{eq:wDE_background_expression}
		w_{DE}=-1+\frac{2(1+n\Omega_Q)(1+z)E_{,z}/E-3\Omega_m-4\Omega_r}{3\Omega_{DE}}.
	\end{equation}
	
	Equation~(\ref{eq:wDE_background_expression}) is the preferred expression for the numerical reconstruction of $w_{DE}(z)$. The agreement between Eqs.~(\ref{eq:wDE_continuity_form}) and (\ref{eq:wDE_background_expression}) also provides an independent consistency test of the numerical evolution.
	
	The three familiar regimes are characterized by
	\begin{equation}\label{eq:wDE_regimes}
		w_{DE}>-1\quad{\rm(quintessence\mbox{-}like)},\qquad w_{DE}=-1,\qquad w_{DE}<-1\quad{\rm(phantom\mbox{-}like)}.
	\end{equation}
	
	No particular regime is assumed in advance; its realization follows from the cosmological evolution and the choice of the model parameters.
	
	\subsection{Deceleration Parameter and $Om$ Diagnostic}\label{subsec:q_om}
	
	The deceleration parameter can be expressed in terms of the normalized Hubble function as
	\begin{equation}\label{eq:q_redshift_definition}
		q(z)=-1+(1+z)\frac{E_{,z}}{E}.
	\end{equation}
	
	Using the Ricci-cutoff relation, an algebraically equivalent expression is
	\begin{equation}\label{eq:q_algebraic_expression}
		q(z)=1-\left[\frac{\Omega_{DE}(z)E^\Delta(z)}{\mathcal{C}_{B}}\right]^{2/(2-\Delta)}.
	\end{equation}
	
	Equations~(\ref{eq:q_redshift_definition}) and (\ref{eq:q_algebraic_expression}) provide another numerical consistency check. Accelerated expansion corresponds to $q<0$, whereas $q>0$ characterizes a decelerating phase.
	
	The transition redshift $z_t$ is determined by
	\begin{equation}\label{eq:transition_redshift_definition}
		q(z_t)=0.
	\end{equation}
	
	Using Eq.~(\ref{eq:q_algebraic_expression}), this condition can equivalently be written as
	\begin{equation}\label{eq:transition_redshift_relation}
		\Omega_{DE}(z_t)E^\Delta(z_t)=\mathcal{C}_{B}.
	\end{equation}
	
	The expansion history can also be examined using the $Om$ diagnostic, which provides a useful geometrical test of deviations from the late-time flat $\Lambda$CDM expansion \cite{Sahni2008Om}. For $z\neq0$ it is defined as
	\begin{equation}\label{eq:Om_diagnostic}
		Om(z)=\frac{E^2(z)-1}{(1+z)^3-1}.
	\end{equation}
	
	At $z=0$, Eq.~(\ref{eq:Om_diagnostic}) has the indeterminate form $0/0$ and the corresponding present value, if required, must be understood through the $z\to0$ limit rather than direct substitution. For a spatially flat $\Lambda$CDM universe containing pressureless matter and a cosmological constant, and neglecting radiation at late times, $Om(z)$ remains constant and equal to $\Omega_{m0}$. A nonconstant behavior therefore signals a deviation of the background expansion from this reference model.
	
	For completeness, the total effective equation-of-state parameter can be defined kinematically as
	\begin{equation}\label{eq:weff_definition}
		w_{\rm eff}\equiv-1-\frac{2\dot H}{3H^2}.
	\end{equation}
	
	In terms of the deceleration parameter,
	\begin{equation}\label{eq:weff_q_relation}
		w_{\rm eff}=-1+\frac{2(1+z)}{3}\frac{E_{,z}}{E}=\frac{2q-1}{3}.
	\end{equation}
	
	The quantity $w_{\rm eff}$ characterizes the total expansion history and should not be identified with $w_{DE}$, which refers specifically to the Barrow holographic component.
	
	\subsection{Jerk and Statefinder Diagnostics}\label{subsec:jerk_statefinder}
	
	Higher-order kinematic quantities provide additional information about the evolution of the cosmic expansion. The jerk parameter is defined by
	\begin{equation}\label{eq:jerk_definition}
		j\equiv\frac{\dddot a}{aH^3}.
	\end{equation}
	
	It can be expressed in terms of the deceleration parameter as
	\begin{equation}\label{eq:jerk_q_relation}
		j(z)=q(2q+1)+(1+z)\frac{dq}{dz}.
	\end{equation}
	
	An equivalent expression in terms of $E(z)$ is
	\begin{equation}\label{eq:jerk_E_relation}
		j(z)=1-2(1+z)\frac{E_{,z}}{E}+(1+z)^2\left[\left(\frac{E_{,z}}{E}\right)^2+\frac{E_{,zz}}{E}\right].
	\end{equation}
	
	In the numerical implementation, the jerk is evaluated through Eq.~(\ref{eq:jerk_q_relation}). The derivative $dq/dz$ is obtained analytically by differentiating $\mathcal{D}$ and the positive branch $\mathcal{B}=(\mathcal{D}/\mathcal{C}_{B})^{2/(2-\Delta)}$ along the first-order background solution. Thus, the reported jerk values do not rely on a finite-difference estimate of $E_{,zz}$ or on numerical second differentiation of the Hubble function.
	
	The Statefinder diagnostic is defined by \cite{Sahni2003Statefinder}
	\begin{equation}\label{eq:statefinder_definition}
		r\equiv j,\qquad s\equiv\frac{r-1}{3(q-1/2)}.
	\end{equation}
	
	For the late-time spatially flat matter--$\Lambda$ reference, neglecting radiation, the characteristic Statefinder point is
	\begin{equation}\label{eq:LCDM_statefinder_point}
		(r,s)=(1,0).
	\end{equation}
	
	The evolution of the present model in the $(r,s)$ plane can therefore be compared with this late-time $\Lambda$CDM reference point. The algebraic definition of $s$ is formally undefined at $q=1/2$; consequently, the Statefinder trajectory must be interpreted carefully when the background approaches the matter-dominated regime.
	
	\subsection{Effective Energy Conditions and Adiabatic Background Diagnostic}\label{subsec:energy_stability}
	
	To discuss the background energy conditions, the modified cosmological equations can be rewritten in an effective Einstein-like form by defining
	\begin{equation}\label{eq:effective_density_pressure}
		\rho_{\rm eff}\equiv3H^2,\qquad p_{\rm eff}\equiv-2\dot H-3H^2.
	\end{equation}
	
	These quantities characterize the total effective background and should not be interpreted as the energy density and pressure of an additional fundamental material component. In modified gravity, this distinction is important when interpreting the standard energy conditions \cite{Visser1997EnergyConditions}. The conditions introduced below therefore provide a kinematic characterization of the effective background expansion; they are not independent no-ghost or gradient-stability criteria for the underlying modified-gravity theory.
	
	Using the deceleration parameter, the combinations relevant to the energy conditions are
	\begin{equation}\label{eq:effective_energy_combinations}
		\rho_{\rm eff}+p_{\rm eff}=2H^2(1+q),\qquad \rho_{\rm eff}+3p_{\rm eff}=6H^2q,\qquad \rho_{\rm eff}-p_{\rm eff}=2H^2(2-q).
	\end{equation}
	
	The null energy condition becomes
	\begin{equation}\label{eq:NEC_condition}
		{\rm NEC}:\qquad \rho_{\rm eff}+p_{\rm eff}\geq0\qquad\Longleftrightarrow\qquad q\geq-1.
	\end{equation}
	
	The weak energy condition requires
	\begin{equation}\label{eq:WEC_condition}
		{\rm WEC}:\qquad \rho_{\rm eff}\geq0,\qquad q\geq-1.
	\end{equation}
	
	For the expanding solutions considered here, $\rho_{\rm eff}=3H^2\geq0$ is automatically satisfied.
	
	The strong energy condition reduces to
	\begin{equation}\label{eq:SEC_condition}
		{\rm SEC}:\qquad q\geq0.
	\end{equation}
	
	Hence, accelerated expansion with $q<0$ corresponds to a violation of the effective strong energy condition.
	
	The dominant energy condition yields
	\begin{equation}\label{eq:DEC_condition}
		{\rm DEC}:\qquad -1\leq q\leq2.
	\end{equation}
	
	We finally consider the adiabatic background diagnostic of the Barrow holographic component,
	\begin{equation}\label{eq:ca2_definition}
		c_a^2\equiv\frac{\dot p_{DE}}{\dot\rho_{DE}}.
	\end{equation}
	
	Using $p_{DE}=w_{DE}\rho_{DE}$ and the dark-energy conservation equation gives
	\begin{equation}\label{eq:ca2_w_relation}
		c_a^2=w_{DE}-\frac{\dot w_{DE}}{3H(1+w_{DE})}.
	\end{equation}
	
	In terms of redshift,
	\begin{equation}\label{eq:ca2_redshift}
		c_a^2=w_{DE}+\frac{1+z}{3(1+w_{DE})}\frac{dw_{DE}}{dz}.
	\end{equation}
	
	The quantity $c_a^2$ is an adiabatic background diagnostic and should not be identified with the physical rest-frame propagation speed $c_s^2$ of scalar perturbations in the full modified-gravity--holographic system. In general, $c_a^2\neq c_s^2$. A complete perturbative stability analysis would require the linear perturbation equations together with the corresponding kinetic and gradient terms. At an exact phantom-divide crossing, $w_{DE}=-1$ and the separate continuity equation gives $\dot\rho_{DE}=0$; consequently, both Eq.~(\ref{eq:ca2_redshift}) and the ratio in Eq.~(\ref{eq:ca2_definition}) can become formally undefined or divergent. This behavior is a property of the derived background ratio and does not by itself imply a gradient instability. The background solution may remain smooth through the crossing, while a physical sound-speed and stability assessment requires the perturbation sector, which is beyond the scope of the present work.
	
	The equations obtained in this section form the complete background framework used in the subsequent numerical analysis. Solving Eq.~(\ref{eq:E_redshift_ODE}) determines $E(z)$, after which the density parameters, dark-energy equation of state, deceleration parameter, transition redshift, $Om$ diagnostic, jerk and Statefinder parameters, effective equation of state, effective energy-condition indicators, and the adiabatic background diagnostic follow from the relations derived above. The dependence of these quantities on the model parameters, with particular emphasis on the $(n,\Delta)$ plane, will be investigated in the next section together with the comparison with observational Hubble data.
	
	\section{Parameter Sensitivity and Comparison with Cosmic-Chronometer Hubble Data}
	\label{sec:parameter_observational_analysis}
	
	The background framework developed in Sec.~\ref{sec:cosmological_dynamics} provides a closed description of the late-time evolution in terms of the nonlinear power-law correction to the non-metricity sector and the Barrow holographic component. We now investigate how the two characteristic deformation parameters, $n$ and $\Delta$, modify the expansion history and its associated kinematic quantities. The analysis is organized around a common fiducial cosmology, allowing the individual effects of the power-law and Barrow sectors to be isolated before their combined dependence is examined in the $(n,\Delta)$ plane.
	
	The resulting background solutions are subsequently confronted with direct spectroscopic cosmic-chronometer measurements of $H(z)$. The observational comparison is performed for the representative parameter choices adopted in the sensitivity analysis, with the remaining background quantities held fixed. Accordingly, the resulting Hubble-rate statistic is used as a conditional measure of agreement with the cosmic-chronometer data rather than as a global parameter fit or a determination of confidence regions. This strategy keeps the observational analysis directly tied to the physical trends exhibited by the theoretical solutions while avoiding parameter inferences that are not supported by the present numerical setup.
	
	\subsection{Background Evolution and Parameter Sensitivity}
	\label{subsec:parameter_sensitivity}
	
	The two parameters $n$ and $\Delta$ encode physically distinct modifications of the late-time cosmological background. The index $n$ controls the nonlinear power-law contribution $\alpha Q^n$ in the gravitational sector, whereas $\Delta$ characterizes the Barrow deformation entering the holographic energy density through the Ricci infrared cutoff. To disentangle these effects, we first study each parameter separately around a common fiducial cosmology and subsequently examine their joint dependence in the $(n,\Delta)$ plane.
	
	For the standard background sector, we adopt the Planck 2018 base-$\Lambda$CDM values of $H_0$ and $\Omega_{m0}$ \cite{Planck2020Parameters}. The radiation density is computed from the standard thermal content using $T_{\rm CMB}=2.7255\,{\rm K}$ and $N_{\rm eff}=3.046$. The present amplitude of the nonlinear geometric contribution is fixed at the small benchmark value $\Omega_{Q0}=0.02$. This choice is not interpreted as an observational constraint; rather, it allows the effect of the power-law index to be studied while keeping the present magnitude of the geometric sector under control.
	
	Since the dimensionless power-law amplitude satisfies Eq.~(\ref{eq:OmegaQ0_definition}), fixing $\Omega_{Q0}$ implies
	\begin{equation}\label{eq:alpha_bar_fixed_OmegaQ0}
		\bar{\alpha}(n)=\frac{\Omega_{Q0}}{2n-1}.
	\end{equation}
	Accordingly, $\bar{\alpha}$ is recomputed whenever $n$ is varied. This parametrization separates, as far as possible at the background level, the redshift dependence introduced by $n$ from a simultaneous variation of the present geometric amplitude. The value $n=1/2$ is not included in the fixed-$\Omega_{Q0}$ scans because Eq.~(\ref{eq:alpha_bar_fixed_OmegaQ0}) cannot realize a finite nonzero $\Omega_{Q0}$ at that point. This reflects the background-degenerate character of the $n=1/2$ branch discussed in Sec.~\ref{sec:fqlm_powerlaw}, rather than a singularity of the cosmological evolution itself.
	
	For the reference power-law index, we choose $n_{\rm ref}=-1/3$. This value is motivated by observational analyses of closely related power-law $f(Q)$ cosmologies, which favor values in the vicinity of $n\simeq-1/3$ \cite{Mandal2023PowerLawfQ}. We emphasize that this motivation is used only to define a representative benchmark: the present framework contains an additional Barrow holographic sector, and constraints derived in a pure power-law $f(Q)$ model cannot be transferred directly to the current theory.
	
	A small Barrow deformation, $\Delta_{\rm ref}=0.01$, is adopted as the reference value. This choice is motivated by the tendency toward values close to the undeformed limit found in a related Barrow holographic model with a local Granda--Oliveros infrared cutoff \cite{Oliveros2022BarrowGO}. In particular, the corresponding analysis found a best-fit deformation consistent with $\Delta=0$ and constrained to be very small at the quoted confidence level. The value adopted here is therefore used only as a mild-deformation benchmark and not as an observational determination of $\Delta$ for the present model.
	
	The holographic normalization $\mathcal{C}_B$ is fixed internally rather than imported from another dark-energy model. At the present epoch, the modified closure relation gives
	\begin{equation}\label{eq:OmegaDE0_fiducial_relation}
		\Omega_{DE0}=1-\Omega_{m0}-\Omega_{r0}+\Omega_{Q0}.
	\end{equation}
	We normalize the fiducial background to the present acceleration scale of the corresponding Planck-like flat-$\Lambda$CDM cosmology, for which $q_{0,{\rm ref}}\simeq-0.5273$. Using the exact present-day Barrow--Ricci relation,
	\begin{equation}\label{eq:CB_present_normalization}
		\mathcal{C}_B=\frac{\Omega_{DE0}}{(1-q_0)^{1-\Delta/2}},
	\end{equation}
	then fixes the reference value of $\mathcal{C}_B$. This procedure is a normalization prescription within the present model and does not constitute an additional fit to observational data.
	
	The principal fiducial quantities entering the numerical analysis are summarized in Table~\ref{tab:fiducial}. The listed value of $q_{0,{\rm ref}}$ is the normalization target used to calibrate $\mathcal{C}_{B,{\rm ref}}$ through Eq.~(\ref{eq:CB_present_normalization}); it is not an additional parameter inferred from the data.
	
	\begin{table*}[t]
		\centering
		\caption{Principal fiducial quantities adopted throughout the numerical background analysis. The Hubble constant is given in units of ${\rm km\,s^{-1}\,Mpc^{-1}}$. The value $q_{0,{\rm ref}}$ is used only to calibrate $\mathcal{C}_{B,{\rm ref}}$ through Eq.~(\ref{eq:CB_present_normalization}) and is not an independent observational fit.}
		\label{tab:fiducial}
		\scriptsize
		\begin{ruledtabular}
			\begin{tabular}{ccccccccc}
				$H_0$ & $\Omega_{m0}$ & $\Omega_{r0}$ & $\Omega_{Q0}$ & $n_{\rm ref}$ & $\Delta_{\rm ref}$ & $\bar{\alpha}_{\rm ref}$ & $q_{0,{\rm ref}}$ & $\mathcal{C}_{B,{\rm ref}}$ \\
				\hline
				67.4 & 0.315 & $9.1948\times10^{-5}$ & 0.02 & $-1/3$ & 0.01 & $-0.0120$ & $-0.527316$ & 0.462512
			\end{tabular}
		\end{ruledtabular}
	\end{table*}
	
	To probe the power-law dependence, we consider
	$n\in\{-2/3,-1/3,0,1/3,4/3\}$ while keeping the Barrow sector at its fiducial value and preserving the same present geometric amplitude. The case $n=0$ is retained as the cosmological-constant-like limit of the power-law correction discussed in Sec.~\ref{sec:fqlm_powerlaw}. Conversely, the Barrow sensitivity is explored with
	$\Delta\in\{0,0.01,0.04,0.07,0.10\}$ at fixed $n=n_{\rm ref}$. The value $\Delta=0$ provides the undeformed holographic reference, allowing the effects generated specifically by the Barrow correction to be followed continuously.
	
	For the numerical integration, the background evolution is implemented in the e-fold variable $x=\ln a=-\ln(1+z)$, even though the analytical equations are presented in redshift form in Sec.~\ref{sec:cosmological_dynamics}. The first-order system is integrated separately toward the past and the future with the DOP853 algorithm, using relative and absolute tolerances $10^{-9}$ and $10^{-11}$, respectively. The maximum e-fold step is set to $0.02$ on the past branch and $0.005$ on the future branch. Event conditions terminate a trajectory if $E$ or $\mathcal{D}$ approaches the boundary of the physical domain or if either $f_Q$ or $1+n\Omega_Q$ approaches its regularity threshold. Higher-order diagnostics are reconstructed from the analytical background relations; in particular, the jerk is evaluated from the analytical derivative $dq/dz$ rather than from a numerical second derivative of $E(z)$. For the cosmic-chronometer calculation below, the background solution is evaluated directly at the 36 observed redshifts rather than interpolated from the plotting grid.
	
	For each parameter choice, the background equation Eq.~(\ref{eq:E_redshift_ODE}) is integrated from the present normalization $E(0)=1$ toward both positive and negative redshift. Only regular solutions satisfying
	\begin{equation}\label{eq:parameter_analysis_physical_conditions}
		E>0,\qquad \mathcal{B}>0,\qquad \mathcal{D}>0,\qquad f_Q\neq0,\qquad 1+n\Omega_Q\neq0
	\end{equation}
	throughout the relevant numerical interval are retained. These conditions enforce the positive real Ricci-cutoff branch and exclude the background poles associated with $f_Q=0$ and $1+n\Omega_Q=0$ within the numerical domain considered.
	
	Several independent checks were performed before extracting the cosmological observables. In particular, the numerical solutions were verified against the modified Friedmann closure relation and the Barrow holographic identity, while the algebraic and differential reconstructions of the deceleration parameter were found to agree to numerical precision. The equation of state reconstructed from the background field equations was likewise checked against the independently conserved dark-energy continuity equation. These tests ensure that the parameter dependence discussed below reflects the cosmological dynamics of the model rather than numerical inconsistencies of the integration procedure.
	
	\subsubsection{Normalized Expansion History}
	\label{subsubsec:normalized_hubble_rate}
	
\begin{figure*}[t]
		\centering
		\includegraphics[width=0.96\textwidth]{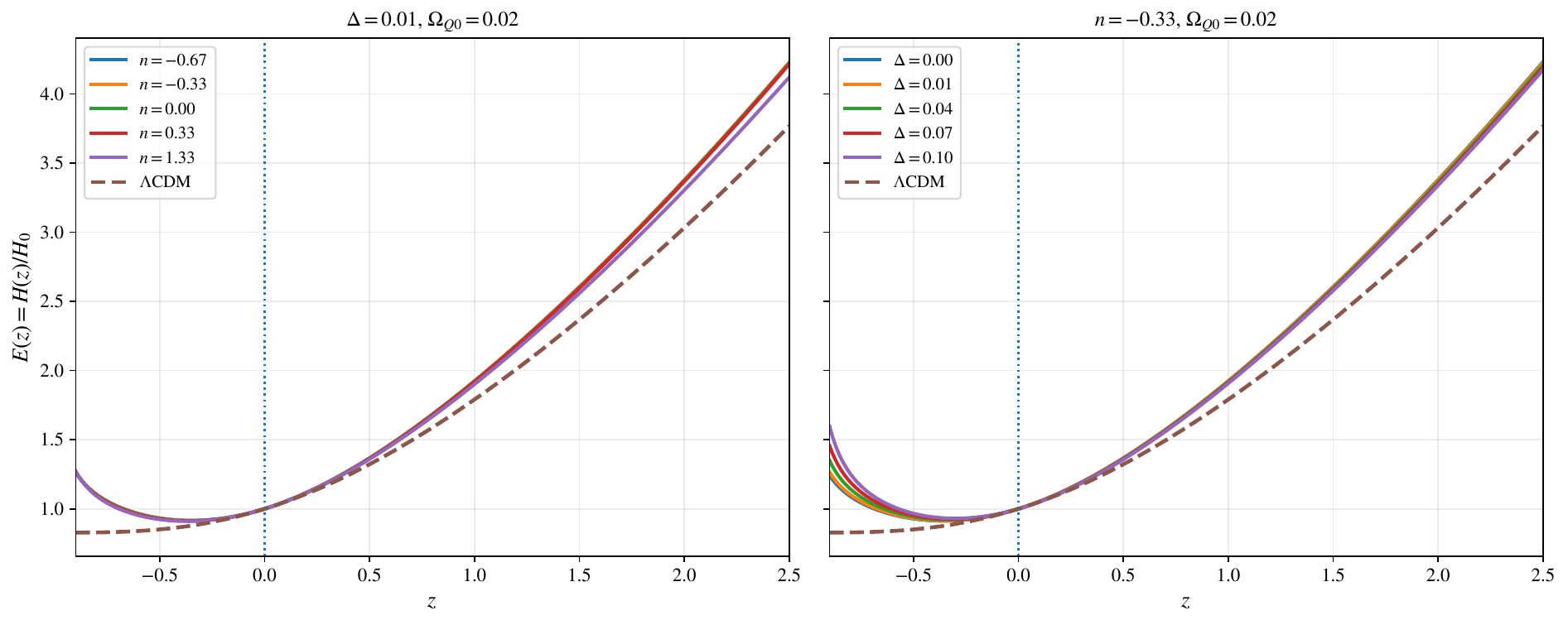}
		\caption{Normalized expansion history $E(z)=H(z)/H_0$ over the interval $-0.9\leq z\leq2.5$. The left panel shows the dependence on the power-law index $n$, whereas the right panel displays the effect of the Barrow deformation parameter $\Delta$. The remaining parameters are fixed to the fiducial values listed in Table~\ref{tab:fiducial}. The dashed curve denotes the corresponding flat-$\Lambda$CDM reference, and the vertical dotted line marks the present epoch $z=0$.}
		\label{fig:E_parameter_sensitivity}
	\end{figure*}

	We begin with the normalized Hubble rate $E(z)=H(z)/H_0$, obtained by integrating Eq.~(\ref{eq:E_redshift_ODE}) from the present condition $E(0)=1$. Figure~\ref{fig:E_parameter_sensitivity} compares the resulting expansion histories when the power-law index $n$ and the Barrow deformation $\Delta$ are varied separately. In each panel, only the indicated parameter is changed, while the remaining quantities are fixed according to Table~\ref{tab:fiducial}. The corresponding flat-$\Lambda$CDM evolution, evaluated with the same standard background parameters, is shown for reference.
	
	All trajectories intersect at $E(0)=1$ by construction and evolve smoothly across the displayed past and future intervals. At positive redshift, the representative solutions remain relatively close to one another, although their departure from the $\Lambda$CDM reference becomes more visible toward larger $z$. The comparison therefore indicates that both deformation parameters modify the late-time expansion history without producing abrupt changes in the background evolution over the range considered.
	
	The dependence on $n$ is comparatively weak. Holding the present geometric amplitude fixed while changing the power-law index mainly modifies the redshift dependence of the nonlinear $Q^n$ contribution. Consequently, the corresponding expansion histories remain tightly grouped. Even at the future endpoint $z=-0.9$, the displayed $n$ models span only
	\[
	1.265\lesssim E(-0.9)\lesssim1.271,
	\]
	showing that the normalized future expansion is only mildly sensitive to $n$ within the representative range adopted here.
	
	The Barrow deformation produces a considerably stronger effect toward the future. While the $\Delta$ trajectories remain close over much of the positive-redshift interval, their separation grows rapidly for $z<0$. In particular,
	\[
	E(-0.9)\simeq1.241
	\]
	for the undeformed case, whereas
	\[
	E(-0.9)\simeq1.597
	\]
	for $\Delta=0.10$. The influence of the Barrow correction is therefore progressively amplified as the cosmology evolves beyond the present epoch, in marked contrast with the much weaker future sensitivity to the power-law index.
	
	Figure~\ref{fig:E_parameter_sensitivity} thus provides the first indication that $n$ and $\Delta$ play different dynamical roles. The power-law index primarily induces modest changes in the shape of the expansion history, whereas the Barrow deformation has a stronger impact on the future background evolution. The relative contributions of the matter, holographic, and geometric sectors underlying this behavior are examined next.
	
	\subsubsection{Density-Parameter Evolution}
	\label{subsubsec:density_parameter_evolution}
	
\begin{figure}[t]
		\centering
		\includegraphics[width=0.82\columnwidth]{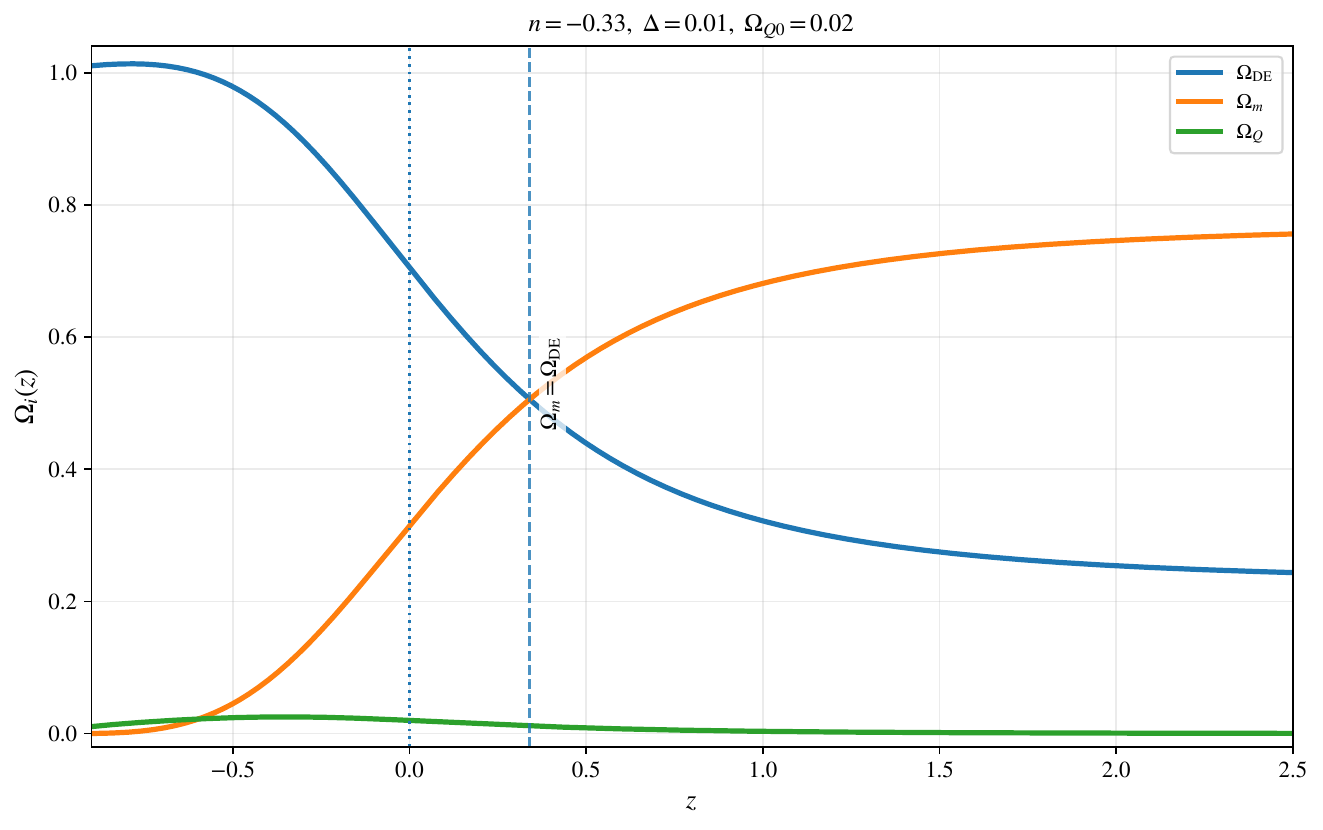}
		\caption{Evolution of the matter density parameter $\Omega_m$, the Barrow holographic dark-energy density parameter $\Omega_{DE}$, and the effective geometric contribution $\Omega_Q$ for the fiducial cosmology. The vertical dotted line marks the present epoch, while the vertical dashed line indicates the matter--dark-energy equality epoch defined by $\Omega_m=\Omega_{DE}$. Radiation is included in the numerical evolution but is omitted from the figure for clarity.}
		\label{fig:density_parameter_evolution}
	\end{figure}

	The origin of the different expansion histories can be understood more directly from the evolution of the individual density components. Figure~\ref{fig:density_parameter_evolution} shows $\Omega_m(z)$, $\Omega_{DE}(z)$, and the effective geometric contribution $\Omega_Q(z)$ for the fiducial model of Table~\ref{tab:fiducial}. The density parameters satisfy the modified closure relation of Eq.~(\ref{eq:powerlaw_closure_relation}), while the geometric sector evolves according to Eq.~(\ref{eq:OmegaQ_E_relation}). Radiation is retained throughout the numerical integration but is omitted from the figure because its contribution is negligible over the late-time interval displayed.
	
	The evolution exhibits the expected transition from matter domination at positive redshift to holographic dark-energy domination at late times. The matter and dark-energy contributions become equal at approximately $z\simeq0.34$. This epoch is distinct from the onset of accelerated expansion: the fiducial solution reaches $q=0$ at the higher redshift $z_t\simeq0.442$, corresponding to an earlier cosmic time than matter--dark-energy equality, as discussed below. The distinction is physically relevant because equality of two density components does not by itself determine the sign of the cosmic acceleration.
	
	Toward higher redshift, $\Omega_m$ increases and rapidly becomes the dominant component, whereas $\Omega_{DE}$ decreases. The geometric contribution remains subdominant and is progressively suppressed toward the past. This behavior follows from $\Omega_Q\propto E^{2(n-1)}$ and the negative fiducial value of $n$: as $E(z)$ grows toward larger redshift, the relative contribution of the nonlinear geometric sector decreases. The resulting hierarchy is therefore compatible with a matter-dominated late-time past while allowing the modified sector to become dynamically relevant at lower redshift.
	
	The future evolution shows the complementary behavior. As $z$ becomes negative, the matter contribution rapidly dilutes and the Barrow holographic sector dominates the energy budget. The figure also shows that $\Omega_{DE}$ can become slightly larger than unity in the future. This is fully consistent with the modified closure relation,
	$\Omega_m+\Omega_r+\Omega_{DE}-\Omega_Q=1$:
	a positive geometric contribution allows $\Omega_{DE}>1$ without violating the Friedmann constraint. Hence, the apparent excess over unity is a consequence of the modified bookkeeping of the background components rather than a failure of normalization.
	
	Figure~\ref{fig:density_parameter_evolution} therefore confirms a regular sequence from matter domination to a holographic-dark-energy-dominated future, with the nonlinear $Q^n$ contribution remaining subdominant but nonvanishing throughout the late-time evolution. The dynamical nature of the dominant holographic component is examined next through its equation-of-state parameter.
	
	\subsubsection{Dark-Energy Equation of State}
	\label{subsubsec:wde_evolution}
	
\begin{figure}[t]
		\centering
		\includegraphics[width=0.96\textwidth]{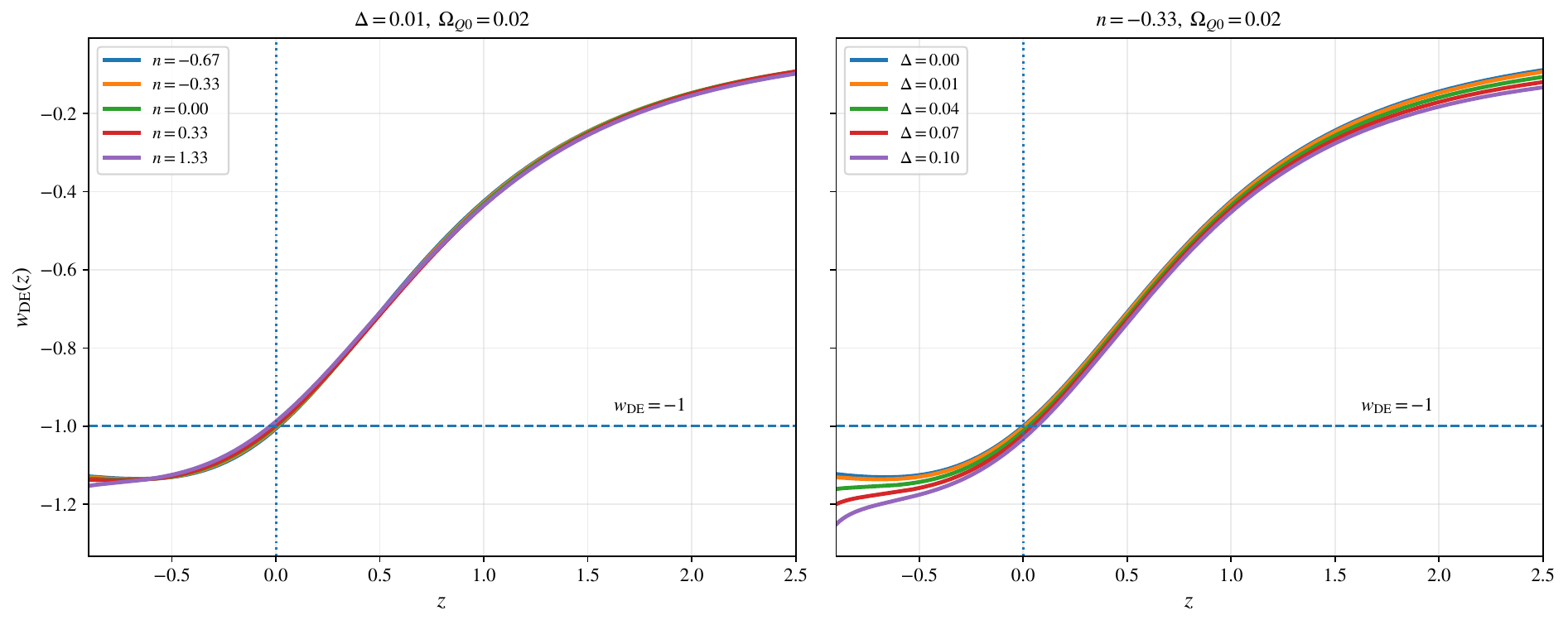}
		\caption{Evolution of the Barrow holographic dark-energy equation-of-state parameter $w_{DE}(z)$. The left panel shows the dependence on the power-law index $n$, whereas the right panel displays the effect of the Barrow deformation parameter $\Delta$. The remaining parameters are fixed to the fiducial values listed in Table~\ref{tab:fiducial}. The horizontal dashed line marks the phantom divide $w_{DE}=-1$, and the vertical dotted line denotes the present epoch $z=0$.}
		\label{fig:wde_parameter_sensitivity}
	\end{figure}

	The dynamical character of the Barrow holographic component is most directly reflected in its effective equation-of-state parameter. Using Eq.~(\ref{eq:wDE_background_expression}), we reconstruct $w_{DE}(z)$ from the numerical background solutions. Figure~\ref{fig:wde_parameter_sensitivity} shows its evolution when $n$ and $\Delta$ are varied separately, with the remaining parameters fixed according to Table~\ref{tab:fiducial}. The horizontal dashed line marks the phantom divide $w_{DE}=-1$, while the vertical dotted line denotes the present epoch.
	
	The left panel reveals a mild but systematic dependence on the power-law index. Across the representative $n$ values, the present equation of state changes from approximately $w_{DE,0}=-1.006$ at $n=-2/3$ to $w_{DE,0}=-0.988$ at $n=4/3$. Increasing $n$ therefore shifts the present cosmology from the slightly phantom side of the $w_{DE}=-1$ boundary toward a mildly quintessence-like regime. The $n=0$ trajectory lies essentially on the cosmological-constant boundary at the present epoch, with $w_{DE,0}\simeq-1$. This relatively small variation is consistent with the weak $n$ dependence of the normalized expansion history found in Fig.~\ref{fig:E_parameter_sensitivity}.
	
	The effect of $n$ is nevertheless genuinely dynamical rather than a constant offset in $w_{DE}$. The trajectories evolve differently away from the present epoch and all of the representative models enter the phantom regime toward the future. At $z=-0.9$, they span approximately $-1.152\lesssim w_{DE}\lesssim-1.128$. Thus, a model that is slightly quintessence-like today can still evolve toward a phantom future, showing that the power-law index controls the time dependence of the holographic sector rather than simply fixing its present equation of state.
	
	The right panel shows a stronger and more uniform sensitivity to the Barrow deformation. Increasing $\Delta$ shifts the equation of state systematically toward more negative values. Along the displayed sequence, $w_{DE,0}$ changes from approximately $-1.000$ in the undeformed limit to $-1.032$ for $\Delta=0.10$. The Barrow correction therefore enhances the phantom character of the present dark-energy sector within the adopted normalization.
	
	This effect becomes considerably stronger toward the future. At $z=-0.9$, the equation of state evolves from approximately $w_{DE}=-1.122$ for $\Delta=0$ to $w_{DE}=-1.251$ for $\Delta=0.10$. The much larger separation of the future trajectories mirrors the behavior of $E(z)$ in Fig.~\ref{fig:E_parameter_sensitivity} and indicates that the influence of the Barrow deformation is progressively amplified as the model evolves beyond the present epoch.
	
	Several trajectories cross the phantom divide close to the present time, while the background quantities themselves remain smooth across the crossing. The fiducial solution, for example, has $w_{DE,0}\simeq-1.003$ and crosses $w_{DE}=-1$ very close to $z=0$. This behavior is a regular feature of the background evolution. The associated adiabatic background diagnostic $c_a^2$, whose redshift representation contains the factor $(1+w_{DE})^{-1}$, is treated separately in Appendix~\ref{app:supplementary_background_diagnostics}; its formal behavior near the crossing is not interpreted as a physical perturbative sound-speed or stability criterion.
	
	At larger positive redshift, the representative trajectories evolve toward less negative values of $w_{DE}$ while $\Omega_{DE}$ simultaneously becomes subdominant, as shown in Fig.~\ref{fig:density_parameter_evolution}. The combined evolution therefore consistently describes a transition from a matter-dominated past to a dynamically evolving holographic-dark-energy phase at late times.
	
	Figure~\ref{fig:wde_parameter_sensitivity} highlights a clear distinction between the two model parameters: $n$ mainly controls the detailed time dependence and the location of the present solution relative to the phantom divide, whereas $\Delta$ more directly determines the strength of the phantom behavior and has a substantially larger impact on the future evolution. The corresponding consequences for the acceleration history are examined next.
	
	\subsubsection{Deceleration Parameter and Transition Redshift}
	\label{subsubsec:q_transition}
	
\begin{figure}[t]
		\centering
		\includegraphics[width=0.96\textwidth]{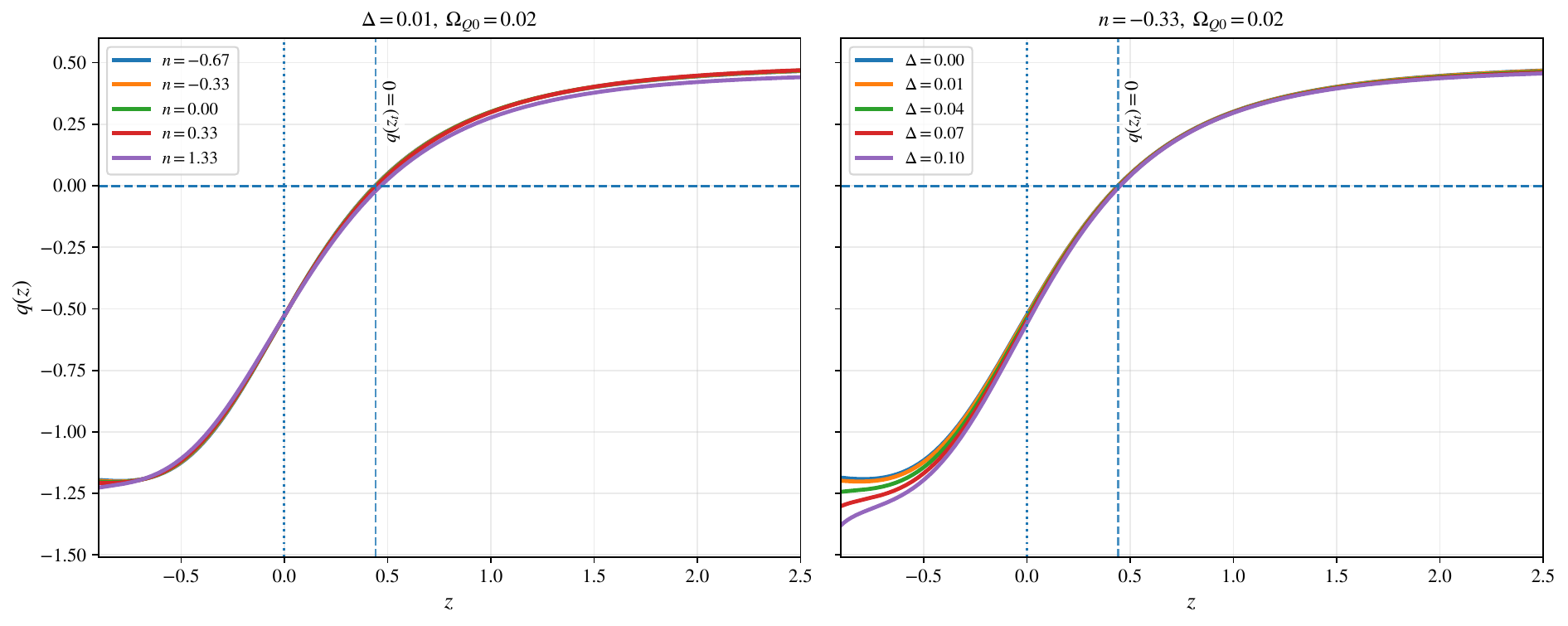}
		\caption{Evolution of the deceleration parameter $q(z)$. The left panel shows the dependence on the power-law index $n$, whereas the right panel displays the effect of the Barrow deformation parameter $\Delta$. The remaining parameters are fixed to the fiducial values listed in Table~\ref{tab:fiducial}. The horizontal dashed line marks $q=0$, the vertical dotted line denotes the present epoch, and the vertical dashed markers indicate the corresponding deceleration-to-acceleration transition redshifts.}
		\label{fig:q_parameter_sensitivity}
	\end{figure}

	The acceleration history is characterized by the deceleration parameter $q(z)$ defined in Eq.~(\ref{eq:q_redshift_definition}). Accelerated expansion corresponds to $q<0$, while $q>0$ describes a decelerating phase. The transition redshift $z_t$ is determined by the condition $q(z_t)=0$, as introduced in Eq.~(\ref{eq:transition_redshift_definition}). Figure~\ref{fig:q_parameter_sensitivity} shows the evolution of $q(z)$ for the same parameter families considered above.
	
	All representative solutions exhibit the same qualitative sequence: a decelerating matter-dominated phase at positive redshift, followed by a transition to accelerated expansion and a progressively stronger accelerated phase toward the future. At larger positive redshift, the trajectories approach the matter-dominated behavior $q\simeq1/2$, consistently with the density evolution shown in Fig.~\ref{fig:density_parameter_evolution}. The transition occurs at relatively recent redshift for all parameter choices considered.
	
	The left panel reveals an important consequence of the fixed-$\Omega_{Q0}$ normalization. Although the power-law index changes the redshift evolution of the geometric sector, all displayed $n$ trajectories share the same present value,
	$q_0=-0.527316$.
	This $n$ independence is not accidental: once $\Omega_{Q0}$, $\Delta$, and the holographic normalization are fixed, the present deceleration parameter is determined by the present-day Barrow--Ricci relation and carries no explicit dependence on $n$. The corresponding two-dimensional behavior is examined separately in Appendix~\ref{app:supplementary_background_diagnostics}.
	
	Away from the present epoch, however, the power-law index leaves a clear imprint on the acceleration history. Across the representative sequence from $n=-2/3$ to $n=4/3$, the transition redshift increases from $z_t\simeq0.440$ to $z_t\simeq0.467$. Larger values of $n$ therefore shift the onset of acceleration to higher redshift, corresponding to an earlier transition in cosmic time, even though the present acceleration is held fixed by the adopted normalization.
	
	The Barrow deformation affects both the present acceleration and the transition epoch. Along the $\Delta$ sequence, $q_0$ decreases from approximately $-0.524$ for $\Delta=0$ to $-0.558$ for $\Delta=0.10$, while the transition redshift increases from $z_t\simeq0.441$ to $z_t\simeq0.452$. Increasing $\Delta$ thus strengthens the present accelerated phase and simultaneously shifts its onset to slightly earlier cosmic times.
	
	The distinction becomes more pronounced on the future branch. At $z=-0.9$, varying $n$ produces only a relatively narrow interval, approximately $-1.23\lesssim q\lesssim-1.20$, whereas the corresponding $\Delta$ sequence spans roughly $-1.38\lesssim q\lesssim-1.19$. This wider separation is consistent with the future behavior of both $E(z)$ and $w_{DE}(z)$ and confirms that, over the representative ranges considered here, the Barrow deformation has the stronger influence on the future late-time acceleration across the displayed interval.
	
	For the fiducial cosmology, the adopted normalization gives $q_0\simeq-0.5273$, while the numerical solution yields $z_t\simeq0.4421$. The latter occurs at a higher redshift than the matter--dark-energy equality epoch, $z\simeq0.34$, found in Fig.~\ref{fig:density_parameter_evolution}. This difference is physically expected: the onset of acceleration is governed by the total effective pressure and the modified gravitational dynamics, rather than solely by equality between the matter and dark-energy density parameters.
	
	Figure~\ref{fig:q_parameter_sensitivity} therefore establishes two complementary effects. The power-law index mainly changes the timing and detailed evolution of the accelerated phase while leaving $q_0$ unchanged under the fixed-$\Omega_{Q0}$ normalization, whereas the Barrow deformation modifies both the present acceleration and its transition history. A complementary geometrical characterization of these departures from the reference $\Lambda$CDM expansion is provided next by the $Om$ diagnostic.
	
	\subsubsection{$Om$ Diagnostic}
	\label{subsubsec:Om_diagnostic_analysis}
	
	The $Om$ diagnostic provides a complementary geometrical characterization of the expansion history using only the normalized Hubble rate. As defined in Eq.~(\ref{eq:Om_diagnostic}), it is constant for a spatially flat $\Lambda$CDM background in the late-time matter--$\Lambda$ limit, where $Om(z)=\Omega_{m0}$. The small radiation contribution retained in our numerical evolution produces only a negligible correction to this reference behavior over the redshift interval considered. A redshift-dependent $Om(z)$ therefore offers a convenient way to visualize departures of the background expansion from the corresponding flat-$\Lambda$CDM cosmology.
	
\begin{figure}[t]
		\centering
		\includegraphics[width=0.96\textwidth]{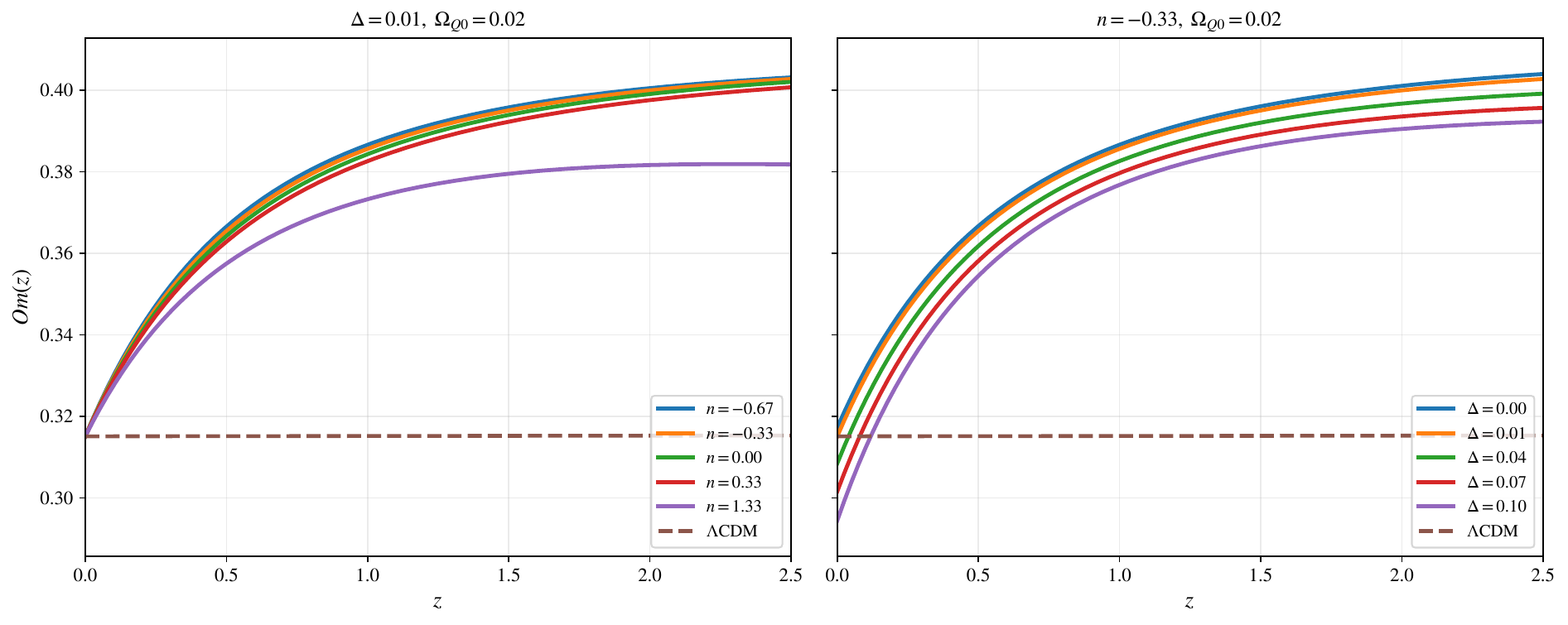}
		\caption{Evolution of the $Om$ diagnostic for the representative parameter families. The left panel shows the dependence on the power-law index $n$, whereas the right panel displays the effect of the Barrow deformation parameter $\Delta$. The remaining parameters are fixed to the fiducial values listed in Table~\ref{tab:fiducial}. The horizontal dashed line denotes the corresponding late-time flat-$\Lambda$CDM reference, $Om\simeq\Omega_{m0}$.}
		\label{fig:Om_parameter_sensitivity}
	\end{figure}

	Figure~\ref{fig:Om_parameter_sensitivity} shows $Om(z)$ for the representative $n$ and $\Delta$ families. Since both the numerator and denominator of the defining expression vanish at $z=0$, the present value is understood through the regular $z\rightarrow0$ limit rather than by direct evaluation of the quotient. The analysis is therefore shown over the positive-redshift interval, where the diagnostic is unambiguous and directly probes differences among the past expansion histories.
	
	The left panel shows that the power-law models develop a clear redshift dependence and depart from the nearly constant $\Lambda$CDM reference. For $n=-2/3$, $-1/3$, $0$, and $1/3$, the trajectories remain relatively close over most of the displayed interval and reach values around $Om\simeq0.40$ at $z=2.5$. The $n=4/3$ trajectory separates more noticeably, reaching approximately $Om\simeq0.382$ at the same redshift. Thus, although the corresponding $E(z)$ curves in Fig.~\ref{fig:E_parameter_sensitivity} are closely grouped, the construction of $Om(z)$ makes their accumulated geometrical differences more apparent.
	
	The Barrow deformation generates a similarly systematic but somewhat more uniform ordering. Increasing $\Delta$ lowers $Om(z)$ across the positive-redshift interval. At $z=2.5$, the diagnostic decreases from approximately $0.404$ in the undeformed case to about $0.392$ for $\Delta=0.10$. The separation among the trajectories grows with redshift, showing that even relatively small changes in the normalized expansion history can lead to distinguishable behavior in this geometrical quantity.
	
	For all representative models, the departure from a nearly constant $Om$ value is therefore a direct manifestation of a background expansion history different from the reference $\Lambda$CDM evolution. The two model parameters affect this departure in different ways: the $n$ dependence remains modest over most of the sampled range but becomes more pronounced for the largest positive value considered, whereas the effect of $\Delta$ is monotonic across the full sequence.
	
	The interpretation of the $Om$ slope requires some care in the present framework. In minimally coupled general-relativistic dark-energy models, the sign of the redshift variation of $Om(z)$ is sometimes used to distinguish effective quintessence-like and phantom-like behavior. Such a correspondence is not model independent and cannot be transferred directly to the present cosmology, where both the nonlinear $Q^n$ contribution and the Barrow holographic sector modify the background dynamics. Here, $Om(z)$ is therefore used strictly as a geometrical measure of departure from the reference expansion history; the physical nature of the dark-energy component is determined independently from $w_{DE}(z)$.
	
	Figure~\ref{fig:Om_parameter_sensitivity} consequently complements the direct Hubble-rate and equation-of-state analyses by exposing differences that are less evident in $E(z)$ alone. To probe the expansion history beyond the Hubble and deceleration levels, we next turn to higher-order kinematic quantities through the jerk and Statefinder parameters.
	
	\subsubsection{Jerk and Statefinder Analysis}
	\label{subsubsec:jerk_statefinder_analysis}
	
	The differences among the expansion histories can be examined beyond the deceleration level by considering higher-order kinematic quantities. The jerk parameter $j(z)$, defined in Eq.~(\ref{eq:jerk_definition}) and related to $q(z)$ through Eq.~(\ref{eq:jerk_q_relation}), is sensitive to the third time derivative of the scale factor and therefore probes features that may remain weakly visible in $E(z)$ or $q(z)$. In the standard spatially flat matter--$\Lambda$ limit, $j=1$; the radiation contribution retained in the numerical calculation produces only a negligible correction to this late-time reference over the redshift range considered here.
	
\begin{figure}[t]
		\centering
		\includegraphics[width=0.96\textwidth]{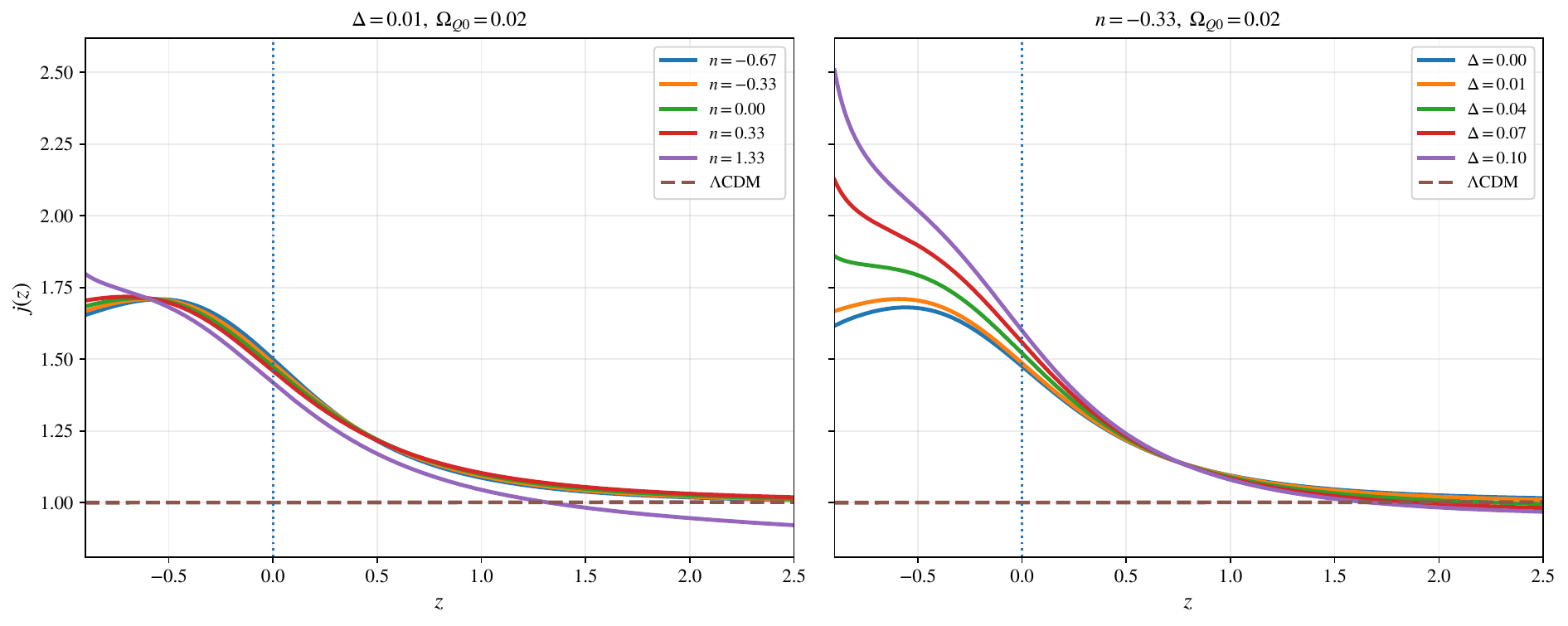}
		\caption{Evolution of the jerk parameter $j(z)$ for the representative parameter families. The left panel shows the dependence on the power-law index $n$, whereas the right panel displays the effect of the Barrow deformation parameter $\Delta$. The remaining parameters are fixed to the fiducial values listed in Table~\ref{tab:fiducial}. The horizontal dashed line denotes the late-time flat-$\Lambda$CDM reference $j=1$, and the vertical dotted line marks the present epoch $z=0$.}
		\label{fig:jerk_parameter_sensitivity}
	\end{figure}

	Figure~\ref{fig:jerk_parameter_sensitivity} shows the jerk evolution for the representative parameter families. The horizontal dashed line denotes the late-time $\Lambda$CDM reference $j=1$, and the vertical dotted line marks the present epoch.
	
	Around the present epoch, all representative trajectories are clearly displaced from the $\Lambda$CDM reference. Along the $n$ sequence, the present jerk decreases monotonically from $j_0\simeq1.500$ for $n=-2/3$ to $j_0\simeq1.418$ for $n=4/3$. The dependence is therefore opposite to that found for the transition redshift: increasing $n$ shifts the acceleration transition to higher redshift while reducing the present value of the jerk. The $n=4/3$ trajectory shows the largest separation from the other members of this family at intermediate and higher redshift and crosses below $j=1$ before the upper end of the plotted interval.
	
	The Barrow deformation produces the opposite trend at the present epoch. Increasing $\Delta$ from $0$ to $0.10$ raises $j_0$ from approximately $1.476$ to $1.599$. The separation among these trajectories becomes particularly pronounced toward the future, consistently with the stronger $\Delta$ dependence already found in $E(z)$, $w_{DE}(z)$, and $q(z)$. Toward positive redshift, however, the jerk curves approach the vicinity of the $\Lambda$CDM reference as the matter contribution becomes increasingly dominant. For the fiducial model, the numerical solution gives $j_0\simeq1.486$.
	
\begin{figure}[t]
		\centering
		\includegraphics[width=0.96\textwidth]{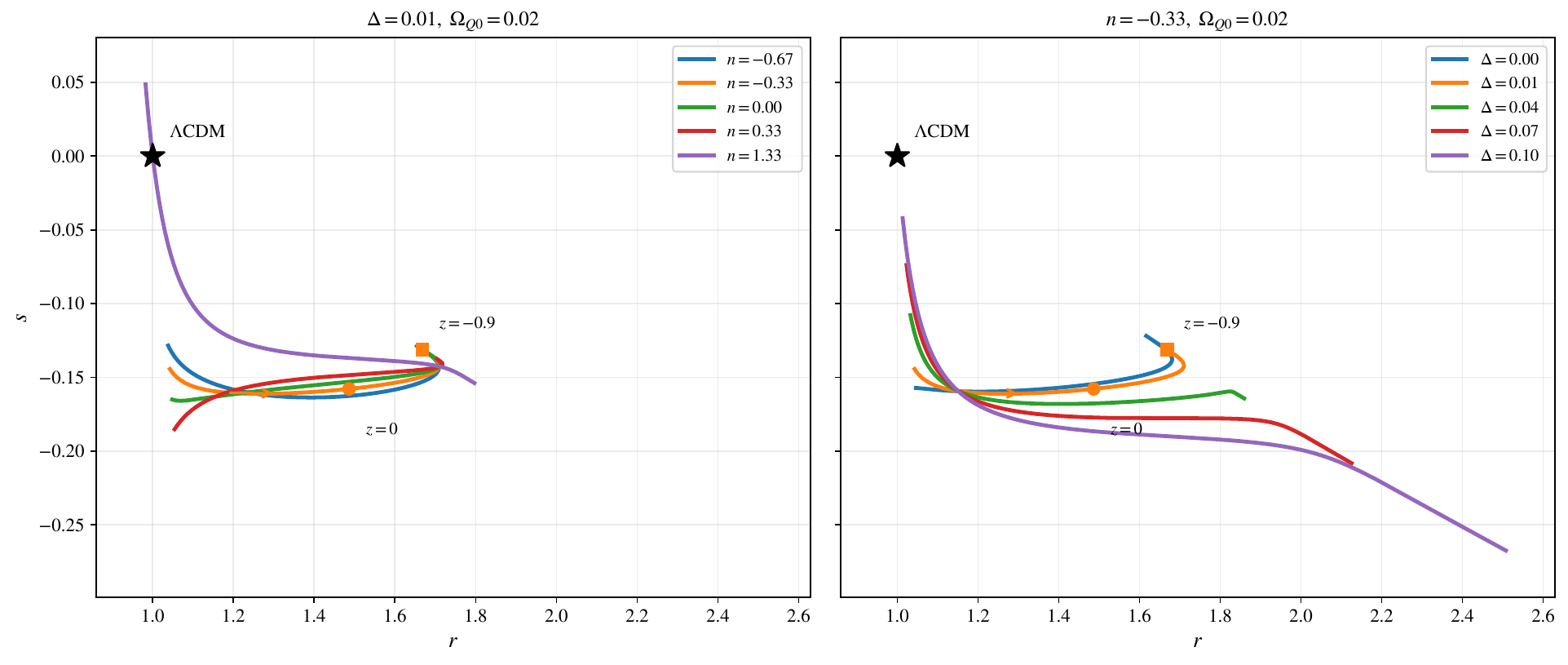}
		\caption{Statefinder trajectories in the $(r,s)$ plane for the representative parameter families. The left panel shows the dependence on the power-law index $n$, whereas the right panel displays the effect of the Barrow deformation parameter $\Delta$. The remaining parameters are fixed to the fiducial values listed in Table~\ref{tab:fiducial}. The black star marks the flat matter--$\Lambda$ reference point $(r,s)=(1,0)$. For the fiducial trajectory, the circle denotes the present epoch $z=0$, while the square marks the future endpoint $z=-0.9$.}
		\label{fig:statefinder_parameter_sensitivity}
	\end{figure}

	A more compact representation of the same higher-order information is provided by the Statefinder pair $(r,s)$ defined in Eq.~(\ref{eq:statefinder_definition}), with $r=j$. The standard flat matter--$\Lambda$ cosmology is associated with the reference point $(r,s)=(1,0)$ given in Eq.~(\ref{eq:LCDM_statefinder_point}). Figure~\ref{fig:statefinder_parameter_sensitivity} traces the trajectories of the representative models in this plane over the same redshift interval used in the preceding background analysis.
	
	Near the present epoch, the representative models occupy the region $r>1$ and $s<0$. The fiducial solution lies at approximately
	\[
	(r_0,s_0)\simeq(1.486,-0.158),
	\]
	providing a higher-order kinematic signature distinct from the reference point. This displacement is consistent with the jerk evolution in Fig.~\ref{fig:jerk_parameter_sensitivity} and demonstrates that models whose normalized Hubble histories remain comparatively close can nevertheless be separated more clearly through higher derivatives of the scale factor.
	
	The power-law index produces a systematic motion of the present Statefinder coordinates. Across the displayed $n$ sequence, $(r_0,s_0)$ changes from approximately $(1.500,-0.162)$ to $(1.418,-0.136)$. Increasing $n$ therefore shifts the present state toward smaller $r$ and less negative $s$. The $n=4/3$ trajectory approaches the vicinity of the $\Lambda$CDM point toward the high-redshift end of the displayed interval. This apparent proximity should not, however, be interpreted as an exact recovery of $\Lambda$CDM, since the Statefinder quantity $s$ becomes increasingly sensitive as $q$ approaches the matter-dominated value $1/2$.
	
	The dependence on $\Delta$ proceeds in the opposite direction. Increasing the Barrow deformation moves the present Statefinder coordinates from approximately $(1.476,-0.155)$ at $\Delta=0$ to $(1.599,-0.189)$ at $\Delta=0.10$. The separation becomes substantially larger toward the future, where the larger-$\Delta$ trajectories extend farther into the region of increasing $r$ and more negative $s$. The Statefinder representation therefore makes the enhanced future sensitivity to the Barrow deformation particularly transparent.
	
	Taken together, Figs.~\ref{fig:jerk_parameter_sensitivity} and \ref{fig:statefinder_parameter_sensitivity} reveal a consistent higher-order distinction between the two deformation sectors. Increasing $n$ lowers the present jerk and moves the Statefinder coordinates toward smaller $r$ and less negative $s$, whereas increasing $\Delta$ produces the reverse trend. These higher-order kinematic quantities consequently complement the expansion-rate, equation-of-state, and deceleration analyses by resolving differences that are less conspicuous at lower derivative order. We next examine how the two parameters act jointly in the $(n,\Delta)$ plane.
	
	\subsubsection{Joint Dependence in the $(n,\Delta)$ Plane}
	\label{subsubsec:joint_n_delta_plane}
	
	The one-parameter sensitivity curves presented above isolate the separate effects of the power-law index $n$ and the Barrow deformation parameter $\Delta$, but they do not fully reveal how the geometric and holographic modifications act simultaneously. To make this interplay explicit, we now examine two-dimensional maps in the $(n,\Delta)$ plane while keeping the remaining background quantities fixed to the fiducial values listed in Table~\ref{tab:fiducial}. At each point of the grid, the dimensionless power-law amplitude $\bar{\alpha}$ is determined from Eq.~(\ref{eq:alpha_bar_fixed_OmegaQ0}) so that the present geometric contribution remains fixed at $\Omega_{Q0}=0.02$.
	
	Among the background observables studied in this work, two are particularly informative for the joint analysis. The first is the present-day dark-energy equation-of-state parameter $w_{{\rm DE},0}$, which directly characterizes the current dynamical state of the Barrow holographic sector. The second is the transition redshift $z_t$, which indicates the epoch at which the cosmic expansion changes from deceleration to acceleration. Taken together, these two quantities provide a compact description of the present and recent background behavior of the model.
	
	\begin{figure}[t]
		\centering
		\includegraphics[width=0.82\columnwidth]{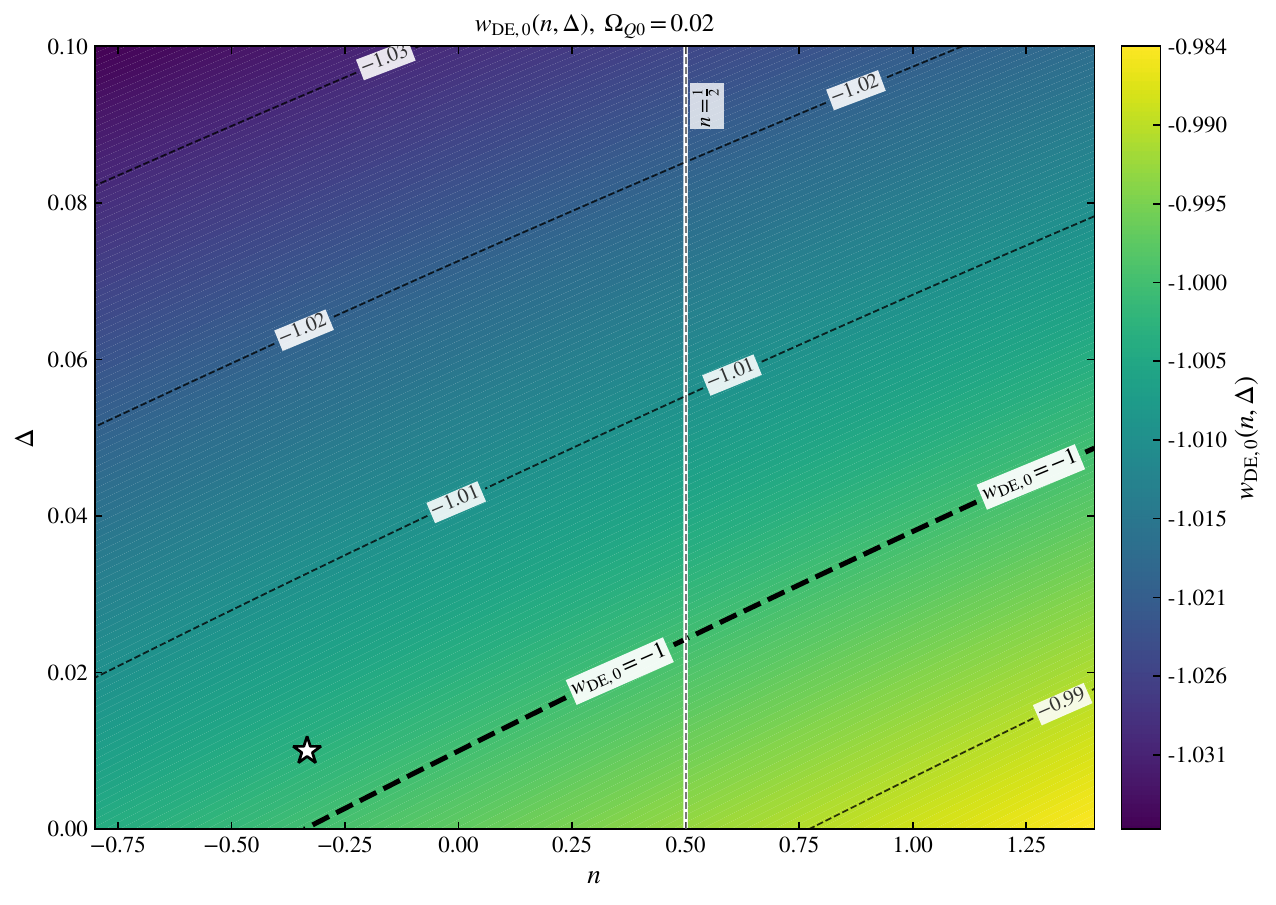}
		\caption{Two-dimensional map of the present-day Barrow holographic dark-energy equation-of-state parameter $w_{{\rm DE},0}$ over $-0.8\leq n\leq1.4$ and $0\leq\Delta\leq0.10$. The remaining parameters are fixed to the fiducial values listed in Table~\ref{tab:fiducial}, and $\bar{\alpha}(n)$ is determined from Eq.~(\ref{eq:alpha_bar_fixed_OmegaQ0}) so that $\Omega_{Q0}=0.02$ remains fixed. The thick black contour marks the phantom-divide boundary $w_{{\rm DE},0}=-1$, while the thinner dashed contours show additional constant-$w_{{\rm DE},0}$ levels. The white star denotes the fiducial point $(n_{\rm ref},\Delta_{\rm ref})=(-1/3,0.01)$. The gray vertical dashed line marks $n=1/2$, where the fixed-nonzero-$\Omega_{Q0}$ parametrization $\bar{\alpha}=\Omega_{Q0}/(2n-1)$ is degenerate; this line is excluded from the valid map and does not represent a cosmological singularity.}
		\label{fig:wde0_parameter_plane}
	\end{figure}
	
	Figure~\ref{fig:wde0_parameter_plane} shows the joint dependence of the present dark-energy equation of state. The map confirms and generalizes the behavior already inferred from the one-parameter scans. For fixed $\Delta$, increasing $n$ shifts $w_{{\rm DE},0}$ toward less negative values, moving the model from the phantom side toward a quintessence-like regime. By contrast, for fixed $n$, increasing $\Delta$ drives $w_{{\rm DE},0}$ toward more negative values and therefore enhances the phantom character of the present dark-energy sector. The two parameters thus act competitively on the present equation of state.
	
	This competition gives rise to a well-defined contour satisfying $w_{{\rm DE},0}=-1$, shown by the thick black contour in Fig.~\ref{fig:wde0_parameter_plane}. This boundary separates the region in which the present Barrow holographic component behaves effectively as quintessence, $w_{{\rm DE},0}>-1$, from the region in which it behaves effectively as phantom dark energy, $w_{{\rm DE},0}<-1$. The fiducial model lies slightly on the phantom side of this boundary, consistently with the one-parameter result $w_{{\rm DE},0}\simeq-1.003$. The existence of this contour is physically useful because it identifies the combinations of geometric and entropic deformations for which the model closely mimics a cosmological-constant-like state at the present epoch.
	
	\begin{figure}[t]
		\centering
		\includegraphics[width=0.82\columnwidth]{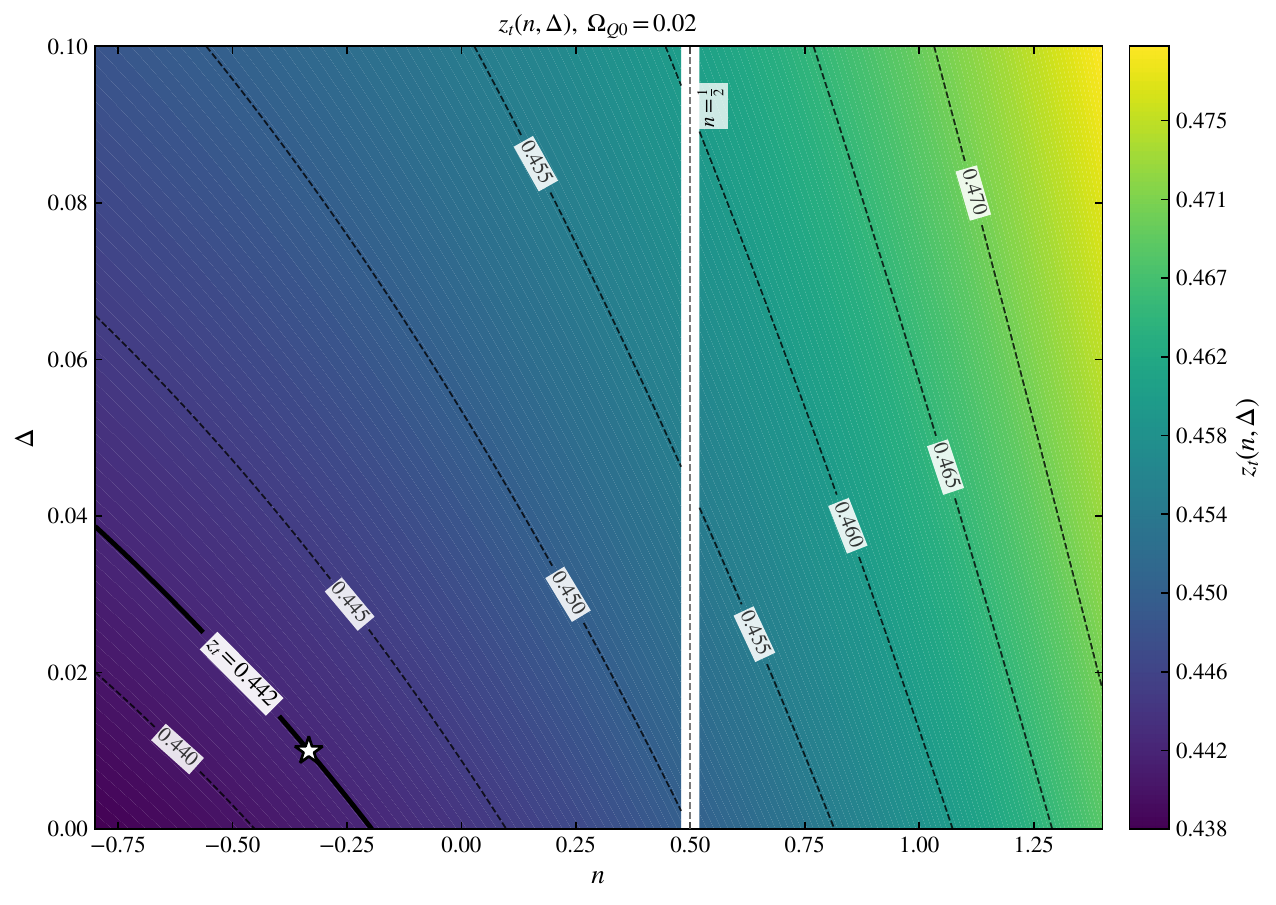}
		\caption{Two-dimensional map of the deceleration-to-acceleration transition redshift $z_t$ over $-0.8\leq n\leq1.4$ and $0\leq\Delta\leq0.10$. The remaining parameters are fixed to the fiducial values listed in Table~\ref{tab:fiducial}, and $\bar{\alpha}(n)$ is determined from Eq.~(\ref{eq:alpha_bar_fixed_OmegaQ0}) so that $\Omega_{Q0}=0.02$ remains fixed. The dashed contours show constant-$z_t$ levels, while the thicker black contour passes through the fiducial value $z_t\simeq0.442$. The white star marks the fiducial point $(n_{\rm ref},\Delta_{\rm ref})=(-1/3,0.01)$. The gray vertical dashed line denotes the parametrization-degenerate value $n=1/2$, which is excluded from the valid map. Larger values of $z_t$ correspond to an earlier onset of cosmic acceleration.}
		\label{fig:zt_parameter_plane}
	\end{figure}
	
	The corresponding map for the transition redshift is displayed in Fig.~\ref{fig:zt_parameter_plane}. In contrast to the competing behavior found for $w_{{\rm DE},0}$, both parameters now act in the same direction: increasing either $n$ or $\Delta$ shifts the transition to larger redshift. Within the fiducial normalization adopted here, stronger geometric or Barrow deformations therefore both lead to an earlier onset of accelerated expansion in cosmic time.
	
	Over the particular parameter intervals sampled here, the total variation of $z_t$ along the $n$ direction is larger than the corresponding variation along the $\Delta$ direction. This statement refers to the finite scanned ranges rather than to a scale-independent comparison of the partial sensitivities, since $n$ and $\Delta$ span different numerical intervals. It is consistent with the one-parameter results, where the change in $z_t$ across the sampled $n$ values is more pronounced than the change induced by the displayed $\Delta$ sequence. The fiducial cosmology is located near $z_t\simeq0.442$, while moving toward larger $n$ and larger $\Delta$ shifts the transition redshift gradually upward.
	
	Taken together, Figs.~\ref{fig:wde0_parameter_plane} and \ref{fig:zt_parameter_plane} provide a compact two-parameter view of the late-time dynamics. The present equation of state reveals a competition between the two deformation sectors: increasing $n$ shifts $w_{{\rm DE},0}$ toward less negative values, whereas increasing $\Delta$ drives it further into the phantom regime. By contrast, their effects on the acceleration-transition redshift are aligned, since larger values of either parameter tend to increase $z_t$. The two observables therefore probe complementary combinations of the geometric and holographic modifications and help disentangle parameter dependences that are not apparent from the individual one-dimensional trajectories alone.
	
	The results of this subsection establish the principal background signatures of the model before observational information is introduced. Across the valid scanned domain, excluding the parametrization-degenerate line $n=1/2$, the retained numerical solutions satisfy the adopted regularity conditions, exhibit a recent transition to accelerated expansion, and display distinct but internally consistent responses to the power-law and Barrow deformations. A supplementary two-dimensional map of the present deceleration parameter, together with the background adiabatic analysis, is presented in Appendix~\ref{app:supplementary_background_diagnostics}. We next assess how the corresponding expansion histories compare with direct spectroscopic measurements of the Hubble rate from cosmic chronometers.
	
	\subsection{Comparison with Cosmic-Chronometer Hubble Data}
	\label{subsec:observational_hubble_data}
	
	The background analysis developed above determines the expansion history for each prescribed set of model parameters. We now examine how these representative trajectories compare with direct measurements of the Hubble rate obtained from cosmic chronometers (CC). The comparison is deliberately conditional: the values of $H_0$, $\Omega_{m0}$, $\Omega_{r0}$, $\Omega_{Q0}$, and the model parameters entering each trajectory are fixed before the observational calculation is performed. No parameter is subsequently varied to improve agreement with the data, and no likelihood optimization, parameter estimation, or confidence-region construction is carried out.
	
	The cosmic-chronometer method determines the expansion rate from the differential aging of passively evolving galaxies. Since
	\begin{equation}\label{eq:cosmic_chronometer_relation}
		H(z)
		=
		-\frac{1}{1+z}\frac{dz}{dt},
	\end{equation}
	measurements of relative galaxy ages at nearby redshifts can be translated into estimates of $H(z)$ without first reconstructing an integrated cosmological distance \cite{JimenezLoeb2002}. This makes CC measurements particularly useful for a direct background-level comparison with the theoretical Hubble histories considered here.
	
	We restrict the observational vector to spectroscopic CC determinations and exclude Hubble-rate measurements inferred from BAO or other distance-scale analyses. The adopted compilation contains 36 measurements over the interval $0.07\leq z\leq1.965$, obtained from differential-age, spectral-index, and full-spectrum-fitting studies \cite{Simon2005,Stern2010,Zhang2014,Moresco2012,Moresco2015,Moresco2016,Ratsimbazafy2017,Jiao2022,Tomasetti2023,Loubser2025,Loubser2025DESI}. A photometric CC determination is not included because the present compilation is intentionally restricted to spectroscopic measurements \cite{Jimenez2023}. The complete observational vector used in the calculation is listed in Table~\ref{tab:cc36_data}.
	
	\begin{table*}[t]
		\centering
		\caption{Spectroscopic cosmic-chronometer measurements of the Hubble parameter adopted in this work. The values of $H(z)$ are given in units of ${\rm km\,s^{-1}\,Mpc^{-1}}$. BAO-derived determinations are not included. For the measurements of Moresco et al.~\cite{Moresco2012,Moresco2016}, the BC03-calibrated values and quoted total uncertainties are used. Statistical and systematic uncertainties are retained separately whenever they are reported separately in the original analyses.}
		\label{tab:cc36_data}
		\scriptsize
		\setlength{\tabcolsep}{3.0pt}
		\renewcommand{\arraystretch}{1.15}
		\begin{tabular}{ccc@{\hspace{0.35cm}}ccc}
			\hline\hline
			$z$ & $H(z)$ & Reference & $z$ & $H(z)$ & Reference \\
			\hline
			0.0700 & $69.0\pm19.6$ & \cite{Zhang2014} & 0.5000 & $72.1\pm33.9_{\rm stat}\pm7.3_{\rm syst}$ & \cite{Loubser2025} \\
			0.1200 & $68.6\pm26.2$ & \cite{Zhang2014} & 0.5929 & $104\pm13$ & \cite{Moresco2012} \\
			0.1700 & $83\pm8$ & \cite{Simon2005} & 0.6700 & $119.45\pm6.39_{\rm stat}\pm16.64_{\rm syst}$ & \cite{Loubser2025DESI} \\
			0.1791 & $75\pm4$ & \cite{Moresco2012} & 0.6797 & $92\pm8$ & \cite{Moresco2012} \\
			0.1993 & $75\pm5$ & \cite{Moresco2012} & 0.7812 & $105\pm12$ & \cite{Moresco2012} \\
			0.2000 & $72.9\pm29.6$ & \cite{Zhang2014} & 0.8000 & $113.1\pm15.1_{\rm stat}{}^{+29.1}_{-11.3}{}_{\rm syst}$ & \cite{Jiao2022} \\
			0.2700 & $77\pm14$ & \cite{Simon2005} & 0.8300 & $108.28\pm10.07_{\rm stat}\pm15.08_{\rm syst}$ & \cite{Loubser2025DESI} \\
			0.2800 & $88.8\pm36.6$ & \cite{Zhang2014} & 0.8754 & $125\pm17$ & \cite{Moresco2012} \\
			0.3519 & $83\pm14$ & \cite{Moresco2012} & 0.8800 & $90\pm40$ & \cite{Stern2010} \\
			0.3802 & $83.0\pm13.5$ & \cite{Moresco2016} & 0.9000 & $117\pm23$ & \cite{Simon2005} \\
			0.4000 & $95\pm17$ & \cite{Simon2005} & 1.0370 & $154\pm20$ & \cite{Moresco2012} \\
			0.4004 & $77.0\pm10.2$ & \cite{Moresco2016} & 1.2600 & $135\pm65$ & \cite{Tomasetti2023} \\
			0.4247 & $87.1\pm11.2$ & \cite{Moresco2016} & 1.3000 & $168\pm17$ & \cite{Simon2005} \\
			0.4497 & $92.8\pm12.9$ & \cite{Moresco2016} & 1.3630 & $160\pm33.6$ & \cite{Moresco2015} \\
			0.4600 & $88.48\pm0.57_{\rm stat}\pm12.32_{\rm syst}$ & \cite{Loubser2025DESI} & 1.4300 & $177\pm18$ & \cite{Simon2005} \\
			0.4700 & $89\pm23_{\rm stat}\pm44_{\rm syst}$ & \cite{Ratsimbazafy2017} & 1.5300 & $140\pm14$ & \cite{Simon2005} \\
			0.4783 & $80.9\pm9.0$ & \cite{Moresco2016} & 1.7500 & $202\pm40$ & \cite{Simon2005} \\
			0.4800 & $97\pm60$ & \cite{Stern2010} & 1.9650 & $186.5\pm50.4$ & \cite{Moresco2015} \\
			\hline\hline
		\end{tabular}
	\end{table*}
	Because the measurements originate from several independent spectroscopic analyses, their reported uncertainty structures are not identical. We therefore retain the information supplied by the original measurements rather than reducing the entire compilation to a single uniform error prescription. For isolated measurements with separately quoted symmetric statistical and systematic uncertainties, the corresponding variance is constructed as
	\begin{equation}\label{eq:CC_single_total_variance}
		\sigma_{H,i}^{2}
		=
		\sigma_{{\rm stat},i}^{2}
		+
		\sigma_{{\rm syst},i}^{2}.
	\end{equation}
	This treatment is used only when no additional covariance information is required for the corresponding measurement.
	
	The determination of Jiao et al.~\cite{Jiao2022} at $z=0.8$ is handled separately because its systematic uncertainty is asymmetric. Instead of replacing the published errors by a symmetric approximation, the lower and upper total uncertainties are retained independently,
	\begin{equation}\label{eq:Jiao_asymmetric_errors}
		\sigma_{J,-}
		=
		\sqrt{15.1^{2}+11.3^{2}},
		\qquad
		\sigma_{J,+}
		=
		\sqrt{15.1^{2}+29.1^{2}}.
	\end{equation}
	The uncertainty associated with this point is selected according to whether the theoretical Hubble rate lies below or above the measured central value,
	\begin{equation}\label{eq:Jiao_piecewise_sigma}
		\sigma_J(H_{\rm th})
		=
		\begin{cases}
			\sigma_{J,-},
			& H_{\rm th}(0.8)<113.1,\\[2mm]
			\sigma_{J,+},
			& H_{\rm th}(0.8)\geq113.1.
		\end{cases}
	\end{equation}
	This prescription retains the directional information carried by the published asymmetric uncertainty rather than introducing an artificial symmetrization.
	
	The three recent DESI-based cosmic-chronometer measurements at $z=0.46$, $0.67$, and $0.83$ also require special treatment because their systematic errors are strongly correlated \cite{Loubser2025DESI}. For the ordered redshift vector $(0.46,0.67,0.83)$, the corresponding covariance block is constructed as
	\begin{equation}\label{eq:DESI_CC_covariance}
		\left(\mathbf{C}_{\rm DESI}\right)_{ij}
		=
		\delta_{ij}\sigma_{{\rm stat},i}^{2}
		+
		\rho_{ij}^{\rm syst}
		\sigma_{{\rm syst},i}
		\sigma_{{\rm syst},j},
	\end{equation}
	with the systematic correlation matrix
	\begin{equation}\label{eq:DESI_CC_correlation}
		\boldsymbol{\rho}_{\rm DESI}^{\rm syst}
		=
		\begin{pmatrix}
			1 & 0.932 & 0.830\\
			0.932 & 1 & 0.776\\
			0.830 & 0.776 & 1
		\end{pmatrix}.
	\end{equation}
	The diagonal elements therefore contain both statistical and systematic variances, while the off-diagonal elements retain the published correlations among the systematic contributions. The three measurements are consequently not treated as independent entries in the comparison.
	
	Apart from this DESI block, no additional cross-survey covariance is introduced unless explicitly supplied for the adopted measurements. In particular, the additional stellar-population-synthesis/model covariance discussed by Moresco et al.~\cite{Moresco2020Covariance} is not reconstructed in the present calculation. This should be regarded as a limitation of the current background-level comparison rather than as an assumption that all residual astrophysical and stellar-population systematics are fundamentally uncorrelated. A more complete covariance treatment would require a dedicated homogeneous reanalysis of the contributing CC samples and is beyond the scope of the present study.
	
	For each representative cosmology considered in Sec.~\ref{subsec:parameter_sensitivity}, the theoretical Hubble rate at the observed redshifts is obtained directly from the already determined background trajectory,
	\begin{equation}\label{eq:H_theory_OHD}
		H_{\rm th}(z) = H_0E(z).
	\end{equation}
	Both $H_0$ and the model parameters defining $E(z)$ are fixed by the corresponding benchmark specification before Eq.~(\ref{eq:H_theory_OHD}) is evaluated. In particular, $H_0$, $n$, $\Delta$, and the remaining background quantities are not adjusted using the CC measurements.
	
	The 35 measurements with symmetric effective uncertainties are collected into a residual vector,
	\begin{equation}\label{eq:Hubble_residual_vector}
		\Delta\mathbf{H}_{\rm s}
		\equiv
		\mathbf{H}_{\rm th}
		-
		\mathbf{H}_{\rm obs},
	\end{equation}
	from which we construct the quadratic discrepancy
	\begin{equation}\label{eq:chi2_CC_symmetric}
		\chi_{\rm s}^{2}
		=
		\Delta\mathbf{H}_{\rm s}^{\,T}
		\mathbf{C}_{\rm CC}^{-1}
		\Delta\mathbf{H}_{\rm s}.
	\end{equation}
	Here $\mathbf{C}_{\rm CC}$ contains the adopted variances of the symmetric-error measurements together with the non-diagonal DESI block defined in Eq.~(\ref{eq:DESI_CC_covariance}).
	
	The asymmetric $z=0.8$ measurement contributes independently through
	\begin{equation}\label{eq:chi2_Jiao}
		\chi_J^{2}
		=
		\frac{\left[H_{\rm th}(0.8)-113.1\right]^{2}}
		{\sigma_J^{2}(H_{\rm th})},
	\end{equation}
	so that the complete conditional statistic for the 36-point compilation is
	\begin{equation}\label{eq:chi2_CC_total}
		\chi_{\rm CC}^{2}
		=
		\chi_{\rm s}^{2}
		+
		\chi_J^{2}.
	\end{equation}
	
	Equation~(\ref{eq:chi2_CC_total}) is evaluated once for each prescribed background trajectory. It is not minimized with respect to any cosmological parameter and is not used to define a best-fitting model, likelihood interval, or statistical confidence region. The resulting value is employed only as a common quantitative measure of the relative agreement between the fixed theoretical trajectories and the same CC data set. Consequently, differences in $\chi_{\rm CC}^{2}$ discussed below should be interpreted within this conditional comparison and not as parameter constraints.
	
	For the graphical comparison, we also use the percentage residual
	\begin{equation}\label{eq:Hubble_percentage_residual}
		\mathcal{R}_H(z_i)
		=
		100\,
		\frac{H_{\rm th}(z_i)-H_{{\rm obs},i}}
		{H_{{\rm obs},i}}.
	\end{equation}
	This quantity provides a dimensionless view of the model--data differences as a function of redshift. It is introduced solely as a visualization diagnostic and does not modify the statistic in Eq.~(\ref{eq:chi2_CC_total}).
	
	The observational construction is therefore kept separate from parameter inference: the CC measurements are used to evaluate a common discrepancy measure for the representative cosmologies already defined by the background analysis, rather than to determine those cosmologies. In the following, we apply this procedure to the same $n$ and $\Delta$ sequences used in the sensitivity study and compare their predicted Hubble histories directly with the 36 spectroscopic CC measurements.
	
	\subsubsection{Conditional Comparison with Cosmic-Chronometer Data}
	\label{subsubsec:conditional_CC_comparison}
	
	We now apply the observational construction described above to the same representative background trajectories used in the parameter-sensitivity analysis. For each trajectory, the theoretical values $H_{\rm th}(z_i)$ are obtained from Eq.~(\ref{eq:H_theory_OHD}) at the redshifts of the 36 spectroscopic cosmic-chronometer measurements listed in Table~\ref{tab:cc36_data}. The corresponding statistic is then evaluated using Eq.~(\ref{eq:chi2_CC_total}), with the correlated DESI covariance block and the asymmetric Jiao et al. contribution treated according to Eqs.~(\ref{eq:DESI_CC_covariance}) and (\ref{eq:chi2_Jiao}), respectively.
	
	For compactness, we denote the value of $\chi_{\rm CC}^{2}$ obtained for a prescribed benchmark trajectory by $\chi^{2}_{H,{\rm cond}}$. This notation does not define a quantity to be minimized over parameter space. Rather, each value is obtained after fixing the cosmological parameters and inserting the resulting background solution into the same CC comparison. In particular, neither $H_0$ nor $n$, $\Delta$, or any of the remaining background parameters is adjusted using the CC measurements.
	
\begin{figure*}[t]
		\centering
		\includegraphics[width=0.98\textwidth]{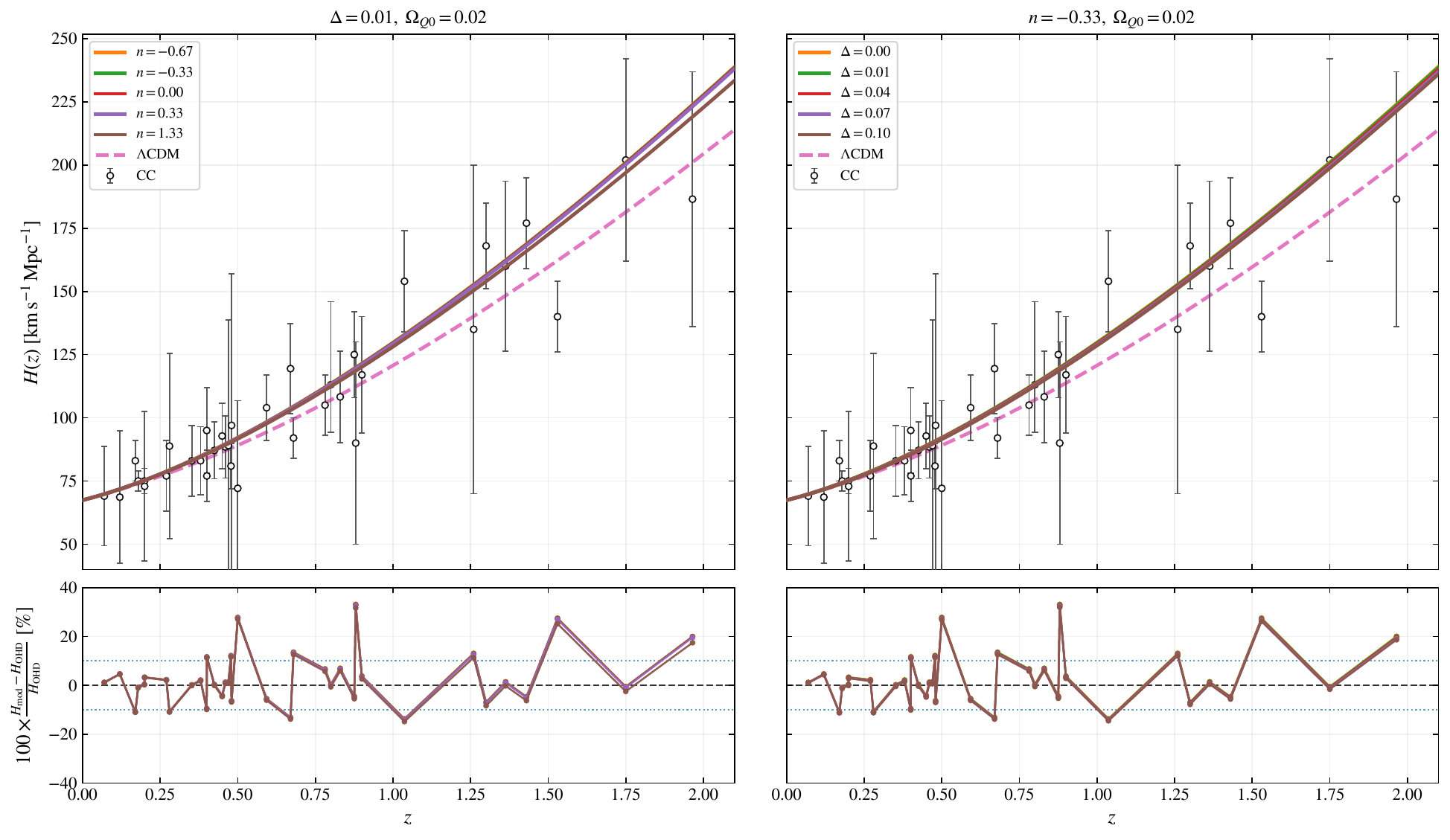}
		\caption{Comparison of the predicted Hubble expansion histories with the 36 spectroscopic cosmic-chronometer measurements listed in Table~\ref{tab:cc36_data}. The left column shows the dependence on the power-law index $n$, while the right column displays the effect of the Barrow deformation parameter $\Delta$; all remaining quantities are fixed according to Table~\ref{tab:fiducial}. The dashed curve in the upper panels denotes the corresponding flat-$\Lambda$CDM reference. The lower panels show the percentage residual $\mathcal{R}_H$ evaluated at the observational redshifts. The horizontal dashed line marks zero residual, the dotted lines indicate $\pm10\%$, and the thin colored lines connecting the residual points are included only as guides to the eye.}
		\label{fig:OHD_parameter_sensitivity}
	\end{figure*}

	Figure~\ref{fig:OHD_parameter_sensitivity} displays the resulting comparison. The upper panels show the predicted Hubble histories together with the CC measurements, while the lower panels display the percentage residual $\mathcal{R}_H$ defined in Eq.~(\ref{eq:Hubble_percentage_residual}). The left column corresponds to the sequence in which $n$ is varied at fixed $\Delta=\Delta_{\rm ref}$, whereas the right column shows the corresponding variation with $\Delta$ at fixed $n=n_{\rm ref}$. All other quantities retain the fiducial values of Table~\ref{tab:fiducial}. The flat-$\Lambda$CDM reference is evaluated with the same standard background normalization and is not fitted independently to the CC sample.
	
	The upper panels show that all representative trajectories follow the overall observed increase of $H(z)$ across the CC redshift range. The different $n$ and $\Delta$ curves remain closely grouped at low and intermediate redshift, where the background predictions are only weakly separated. Their differences become more visible toward the upper part of the observed interval, in agreement with the parameter dependence already identified in the normalized expansion histories of Fig.~\ref{fig:E_parameter_sensitivity}. Over much of the positive-redshift range, the modified-model trajectories also lie somewhat above the corresponding $\Lambda$CDM reference.
	
	The residual panels make the point-by-point differences more transparent. Across the displayed benchmark families, the unweighted root-mean-square percentage residual remains within approximately $10.99\%$--$11.43\%$. Most of the individual residuals have magnitudes below about $20\%$, although several measurements exhibit larger excursions. The largest positive residuals are approximately $32\%$--$33\%$, while the most negative values are around $-14\%$ to $-15\%$. These percentages are intended only as a visualization of the fractional model--data differences. They should not be interpreted as measures of statistical significance, since the observational uncertainties vary substantially among the measurements and the comparison statistic additionally retains the DESI correlations and the asymmetric uncertainty at $z=0.8$.
	
	The conditional statistic provides a covariance-aware summary of the same comparison. For the fiducial trajectory, we obtain $\chi^{2}_{H,{\rm cond}}\simeq21.7701$. Across the displayed $n$ sequence, the corresponding values decrease from approximately $21.8550$ at $n=-2/3$ to $20.6256$ at $n=4/3$. For the $\Delta$ sequence, they decrease from about $21.8790$ at $\Delta=0$ to $20.9178$ at $\Delta=0.10$. The relatively narrow spread of these values is consistent with the substantial overlap among the theoretical Hubble curves in Fig.~\ref{fig:OHD_parameter_sensitivity}.
	
	For comparison, the flat-$\Lambda$CDM trajectory evaluated with the same standard background parameters gives $\chi^{2}_{H,{\rm cond}}\simeq19.7408$. Its smaller value indicates a somewhat lower discrepancy with this particular CC compilation under the common fixed-background prescription. This numerical ordering, however, does not constitute a statistical preference for $\Lambda$CDM over the modified model, because none of the trajectories has been optimized against the data and no likelihood ratio, confidence region, or parameter inference is constructed.
	
	The monotonic decrease of $\chi^{2}_{H,{\rm cond}}$ along the particular $n$ and $\Delta$ sequences shown here must be interpreted with the same caution. The displayed parameter values were selected to probe the background sensitivity of the model rather than to locate extrema of an observational objective function. Consequently, the smaller conditional statistic found toward the upper end of either sequence does not identify a best-fitting value of $n$ or $\Delta$, nor does it imply that the boundaries of the displayed ranges are observationally preferred.
	
	Within these limitations, the CC comparison shows that the representative cosmologies preserve the observed qualitative redshift dependence of the Hubble rate while producing only modest changes in the conditional discrepancy statistic across the displayed parameter families. The purpose of the calculation is therefore to assess the behavior of the preselected background solutions against a common late-time data set, not to determine the cosmological parameters from the data. The associated background quantities and conditional CC statistics are collected systematically in Table~\ref{tab:representative_cosmological_results} in the following numerical summary.
	
	\subsection{Summary of Numerical Results}
	\label{subsec:numerical_summary}
	
	The principal numerical results of the background analysis and the conditional comparison with the cosmic-chronometer sample are summarized in Table~\ref{tab:representative_cosmological_results}. The table collects the present-day deceleration parameter, the dark-energy equation-of-state parameter, the transition redshift, the jerk parameter, and the conditional cosmic-chronometer statistic for the same representative trajectories investigated throughout this section. All parameters that are not explicitly varied in each family are fixed according to the fiducial parameter set given in Table~\ref{tab:fiducial}.
	
	It is important to distinguish derived background quantities from the fiducial normalization prescription. The values of $w_{{\rm DE},0}$, $z_t$, and $j_0$ are derived from the corresponding background solutions. For the fiducial trajectory, however, $q_{0,{\rm ref}}$ is the normalization target used to calibrate $\mathcal{C}_{B,{\rm ref}}$ through Eq.~(\ref{eq:CB_present_normalization}), and the common $q_0$ value along the fixed-$\Delta$, fixed-$\Omega_{Q0}$ $n$ sequence follows analytically from that normalization. By contrast, the variation of $q_0$ along the fixed-$\mathcal{C}_B$ $\Delta$ sequence is a derived response of those trajectories. The quantity $\chi^2_{H,{\rm cond}}$ is included only as a conditional measure of the agreement between these fixed benchmark trajectories and the available cosmic-chronometer measurements; it does not correspond to a likelihood minimization, a best-fit value, or a statistical constraint on the model parameters.

	\begin{table*}[t]
		\caption{Summary of the representative background solutions and their conditional comparison with the cosmic-chronometer data. In the first block, the Barrow deformation parameter is fixed to its fiducial value and the power-law index $n$ is varied. In the second block, $n$ is fixed to its fiducial value and $\Delta$ is varied. All remaining parameters are fixed according to Table~\ref{tab:fiducial}. The quantity $\chi^2_{H,{\rm cond}}$ represents the value obtained from evaluating the cosmic-chronometer statistic for each prescribed trajectory and should not be interpreted as a fitted minimum or a parameter constraint.}
		\label{tab:representative_cosmological_results}
		\begin{ruledtabular}
			\begin{tabular}{cccccccc}
				$n$ & $\Delta$ & $\bar{\alpha}$ & $q_0$ &
				$w_{{\rm DE},0}$ & $z_t$ & $j_0$ &
				$\chi^2_{H,{\rm cond}}$ \\
				\hline
				\multicolumn{8}{c}{$\Delta=0.01,\ \Omega_{Q0}=0.02$} \\
				\hline
				$-\frac{2}{3}$ & $0.01$ & $-0.0086$ &
				$-0.527$ & $-1.006$ & $0.440$ & $1.500$ & $21.855$ \\
				
				$-\frac{1}{3}$ & $0.01$ & $-0.0120$ &
				$-0.527$ & $-1.003$ & $0.442$ & $1.486$ & $21.770$ \\
				
				$0$ & $0.01$ & $-0.0200$ &
				$-0.527$ & $-1.000$ & $0.445$ & $1.473$ & $21.660$ \\
				
				$\frac{1}{3}$ & $0.01$ & $-0.0600$ &
				$-0.527$ & $-0.997$ & $0.449$ & $1.459$ & $21.513$ \\
				
				$\frac{4}{3}$ & $0.01$ & $0.0120$ &
				$-0.527$ & $-0.988$ & $0.467$ & $1.418$ & $20.626$ \\
				
				\hline
				\multicolumn{8}{c}{$n=-1/3,\ \Omega_{Q0}=0.02$} \\
				\hline
				
				$-\frac{1}{3}$ & $0.00$ & $-0.0120$ &
				$-0.524$ & $-1.000$ & $0.441$ & $1.476$ & $21.879$ \\
				
				$-\frac{1}{3}$ & $0.01$ & $-0.0120$ &
				$-0.527$ & $-1.003$ & $0.442$ & $1.486$ & $21.770$ \\
				
				$-\frac{1}{3}$ & $0.04$ & $-0.0120$ &
				$-0.537$ & $-1.012$ & $0.446$ & $1.521$ & $21.460$ \\
				
				$-\frac{1}{3}$ & $0.07$ & $-0.0120$ &
				$-0.548$ & $-1.022$ & $0.449$ & $1.558$ & $21.176$ \\
				
				$-\frac{1}{3}$ & $0.10$ & $-0.0120$ &
				$-0.558$ & $-1.032$ & $0.452$ & $1.599$ & $20.918$ \\
				
			\end{tabular}
		\end{ruledtabular}
	\end{table*}

	The first block of Table~\ref{tab:representative_cosmological_results} illustrates the impact of the power-law index $n$ while keeping the Barrow deformation fixed. Since the present geometric contribution, the Barrow deformation, and the holographic normalization remain fixed within this family, the value of the deceleration parameter is identical to numerical precision for all displayed trajectories, as expected analytically from the present-day Barrow--Ricci relation. In contrast, the present dark-energy equation of state exhibits a smooth evolution from a mildly phantom regime toward a quintessence-like behavior, changing from approximately
	$w_{{\rm DE},0}\simeq -1.006$
	to
	$w_{{\rm DE},0}\simeq -0.988$.
	
	The variation of $n$ also modifies the characteristic time scale of the transition from decelerated to accelerated expansion. The transition redshift increases gradually from approximately
	$z_t\simeq0.440$
	to
	$z_t\simeq0.467$,
	while the jerk parameter decreases from
	$j_0\simeq1.500$
	to
	$j_0\simeq1.418$.
	Therefore, the parameter $n$ mainly affects the detailed dynamical evolution of the dark-energy sector rather than the present normalization of the expansion history.

	The second block demonstrates the distinct role of the Barrow deformation parameter $\Delta$. Increasing $\Delta$ leads to a progressively more negative present deceleration parameter and shifts the equation of state toward a stronger phantom-like regime. Over the displayed range,
	$q_0$
	changes from approximately
	$-0.524$
	to
	$-0.558$,
	whereas
	$w_{{\rm DE},0}$
	moves from a value close to the phantom divide to approximately
	$-1.032$.
	
	At the same time, the transition redshift varies moderately, increasing from
	$z_t\simeq0.441$
	to
	$z_t\simeq0.452$,
	while the jerk parameter increases from approximately
	$1.476$
	to
	$1.599$.
	These trends are consistent with the comparatively stronger influence of $\Delta$ on the late-time acceleration dynamics over the representative parameter ranges examined in the one-dimensional evolutions and parameter-space maps. For the fiducial trajectory, the main background quantities are
	\begin{equation}
		\label{eq:fiducial_summary_results}
		q_0=-0.5273,\qquad
		w_{{\rm DE},0}=-1.0030,\qquad
		z_t=0.4421,\qquad
		j_0=1.4864 .
	\end{equation}
	The corresponding conditional comparison with the cosmic-chronometer data gives
	\begin{equation}
		\label{eq:fiducial_summary_chi2}
		\chi^2_{H,{\rm cond}}=21.7701 .
	\end{equation}
	The identical fiducial values recovered from the two independent parameter scans provide an internal numerical consistency check of the implementation.
	
	Overall, the numerical summary confirms that the two model parameters control different aspects of the late-time cosmological behavior. The power-law index $n$ primarily regulates the evolution of the dark-energy equation of state, the transition redshift, and higher-order kinematic quantities, whereas the Barrow deformation $\Delta$ produces a more direct modification of the acceleration strength and the future evolution of the background over the representative ranges considered here. The conditional cosmic-chronometer statistic varies only modestly among the fixed benchmark trajectories examined in this section. Within the deliberately non-optimized comparison adopted here, this limited spread does not provide a strong separation among those preselected trajectories and should not be interpreted as a parameter constraint.
	
	\section{Conclusions and Outlook}\label{sec:conclusions}
	
	In this work, we have investigated a late-time cosmological framework in which Barrow holographic dark energy with a Ricci infrared cutoff is embedded in the power-law model
	$f(Q,L_m)=Q+\alpha Q^n-2L_m$
	within the broader $f(Q,L_m)$ formulation of symmetric teleparallel gravity. The construction combines two physically distinct deformations: the index $n$ controls the nonlinear dependence of the gravitational sector on the non-metricity scalar, whereas the Barrow parameter $\Delta$ modifies the holographic energy density through the entropy deformation. Since the matter Lagrangian enters linearly and $f_{QL_m}=0$ for the model considered here, the background dynamics contains no explicit mixed matter--geometry coupling. The deviations from the standard symmetric-teleparallel background therefore arise from the nonlinear $Q^n$ contribution, while the Barrow deformation acts through the holographic sector.
	
	A central feature of the formulation is that the cosmological expansion is obtained directly from the modified field equations rather than imposed through a prescribed scale factor or a phenomenological Hubble parametrization. Combining the power-law gravitational equations with the Ricci-cutoff Barrow density reduces the homogeneous dynamics to a closed first-order equation for the normalized Hubble function $E(z)$. The numerical solutions were restricted to the real and regular branch of the theory and were subjected to multiple internal consistency checks, including the modified Friedmann closure, reconstruction of the Barrow density, agreement between the algebraic and differential expressions for the deceleration parameter, and consistency of the dark-energy equation of state with its separate background conservation equation. For fixed $n\neq1/2$, the standard gravitational background is recovered in the limit $\alpha\rightarrow0$, equivalently $\Omega_{Q0}\rightarrow0$ under the parametrization adopted here, while $\Delta\rightarrow0$ continuously reduces the holographic sector to the conventional Ricci holographic form.
	
	The numerical evolution exhibits the expected transition from the matter-dominated portion of the positive-redshift regime to late-time accelerated expansion. For the representative parameter choices, the nonlinear geometric contribution remains subdominant, whereas the Barrow holographic component becomes dynamically important at late times. The two deformation parameters nevertheless leave qualitatively different signatures on the background dynamics. At fixed $\Omega_{Q0}$ and $\Delta$, increasing $n$ shifts $w_{{\rm DE},0}$ toward less negative values, increases the transition redshift, and decreases the present jerk parameter. Under the same fixed-$\Omega_{Q0}$ normalization, $q_0$ has no explicit $n$ dependence. By contrast, at fixed $n$ and fixed $\mathcal{C}_B=\mathcal{C}_{B,\rm ref}$, increasing $\Delta$ makes both $w_{{\rm DE},0}$ and $q_0$ more negative and increases both $z_t$ and $j_0$. The two deformations therefore act in opposite directions on the present dark-energy equation of state, while shifting the onset of cosmic acceleration in the same direction over the representative ranges considered here. The separation among the $\Delta$ trajectories also becomes more pronounced toward the future over the interval explicitly explored in the numerical analysis.
	
	For the fiducial normalization, the reference value
	$q_{0,\rm ref}\simeq-0.5273$
	is used to calibrate the Barrow normalization parameter $\mathcal{C}_{B,\rm ref}$. With this normalization fixed, the resulting fiducial background trajectory yields
	\begin{equation}\label{eq:conclusion_fiducial_predictions}
		w_{{\rm DE},0}\simeq-1.0030,\qquad
		z_t\simeq0.4421,\qquad
		j_0\simeq1.4864.
	\end{equation}
	The present dark-energy equation of state therefore lies only slightly below the phantom divide, and the fiducial trajectory crosses $w_{DE}=-1$ smoothly very close to the present epoch. This crossing does not signal a singularity of the homogeneous background solution.
	
	We have also compared the representative expansion histories with a curated compilation of 36 spectroscopic cosmic-chronometer measurements over $0.07\leq z\leq1.965$. The comparison retains the published covariance among the three DESI cosmic-chronometer measurements and treats the asymmetric uncertainty of the $z=0.8$ measurement without symmetrization. The cosmological parameters are fixed before the observational statistic is evaluated: no minimization, parameter estimation, likelihood reconstruction, posterior inference, or confidence-region analysis is performed. The resulting $\chi^2_{H,\rm cond}$ values therefore provide only conditional discrepancy measures for the prescribed background trajectories relative to a common data set.
	
	For the fiducial trajectory,
	\begin{equation}\label{eq:conclusion_fiducial_chi2}
		\chi^2_{H,\rm cond}\simeq21.7701.
	\end{equation}
	Across the representative modified trajectories, the conditional statistic remains in the comparatively narrow range
	\begin{equation}\label{eq:conclusion_chi2_range}
		20.63\lesssim\chi^2_{H,\rm cond}\lesssim21.88.
	\end{equation}
	The flat-$\Lambda$CDM trajectory evaluated with the same standard background normalization gives the smaller value
	\begin{equation}\label{eq:conclusion_lcdm_chi2}
		\chi^2_{H,\rm cond}(\Lambda{\rm CDM})\simeq19.7408.
	\end{equation}
	Within the deliberately non-optimized comparison performed here, this means only that the reference $\Lambda$CDM trajectory has a smaller conditional discrepancy with the adopted cosmic-chronometer sample than the representative modified trajectories. It does not constitute evidence for statistical preference, exclusion, or parameter constraints. Likewise, the modest variation of $\chi^2_{H,\rm cond}$ among the fixed modified trajectories does not provide a strong separation among those preselected backgrounds. Additional covariance associated with stellar-population synthesis and modeling, beyond the covariance information explicitly implemented in the present calculation, has not been reconstructed and remains a limitation of this background-level comparison \cite{Moresco2020Covariance}.
	
	The supplementary adiabatic analysis further clarifies the scope of the present results. The quantity $c_a^2=\dot p_{DE}/\dot\rho_{DE}$ is used only as a background adiabatic diagnostic. Its redshift representation can become formally undefined or divergent as $w_{DE}\rightarrow-1$, while the homogeneous solution itself remains regular through the crossing. Moreover, $c_a^2$ should not be identified with the physical rest-frame scalar propagation speed $c_s^2$ of the complete modified-gravity--holographic system. Consequently, neither the formal behavior near the phantom divide nor the mildly negative values reached by $c_a^2$ for some representative trajectories provide, by themselves, a no-ghost or gradient-stability criterion.
	
	The present study should therefore be regarded as a systematic background-level characterization of the model rather than a complete assessment of its observational viability. Several extensions are particularly important. First, a genuine parameter-inference analysis should combine the background model with complementary late-time probes, including Type Ia supernovae and baryon acoustic oscillations, together with the cosmic-chronometer sample and its relevant covariance structure. Such an analysis would determine whether the parameter dependences identified here persist when the full background parameter set is varied simultaneously and would permit statistically meaningful parameter constraints and model comparison.
	
	Second, extending the theory to linear cosmological perturbations is essential. Deriving the scalar perturbation equations would allow the physical propagation speed, no-ghost and gradient-stability conditions, the growth of matter perturbations, and the gravitational potentials to be determined consistently. This step is also required before confronting the model with observables such as $f\sigma_8$, weak gravitational lensing, and the full cosmic microwave background anisotropy spectra. A Boltzmann-level implementation would therefore provide a substantially stronger test than is possible from homogeneous expansion data alone and would naturally enable a future joint analysis including CMB information.
	
	A third direction concerns the gravitational sector itself. The power-law model considered here belongs to the $f(Q,L_m)$ framework but has no explicit nonlinear matter--geometry coupling because its dependence on $L_m$ is linear. Extending the analysis to genuinely nonminimal forms with $f_{QL_m}\neq0$ would introduce modified conservation dynamics and allow the interplay among Barrow holography, non-metricity, and explicit matter--geometry coupling to be studied in a more general setting. Such an extension could reveal background and perturbative signatures that are absent in the separable model examined here.
	
	Finally, the stronger sensitivity of the solutions to $\Delta$ toward the future over the interval explored here motivates a dedicated analysis of the asymptotic phase space. The present numerical evolution demonstrates a smooth future branch over the range considered, but it is not sufficient to establish the ultimate cosmic fate. A dynamical-systems treatment, together with an examination of possible future singularities and attractor solutions, would clarify whether the phantom-like trajectories approach a regular asymptotic state or evolve toward qualitatively different late-time regimes.
	
	In summary, the combination of a nonlinear power-law non-metricity sector with Ricci-cutoff Barrow holographic dark energy provides a coherent and dynamically nontrivial background cosmology. The two deformation parameters generate distinct signatures in the dark-energy equation of state, acceleration history, and higher-order kinematics, while the fixed benchmark trajectories reproduce the broad redshift dependence of the cosmic-chronometer measurements with conditional discrepancy statistics that vary only modestly across the representative families. These results provide a well-defined background framework for the perturbative, asymptotic, and multi-probe observational investigations required to determine the full cosmological viability of the model.

	
	\appendix
	\section{Supplementary Background Diagnostics}
	\label{app:supplementary_background_diagnostics}
	
	The main analysis has concentrated on the late-time expansion history, the evolution of the Barrow holographic component, the acceleration transition, higher-order kinematic quantities, and the conditional comparison with cosmic-chronometer measurements. In this Appendix, we present two supplementary background diagnostics that clarify structural features of the numerical solutions without introducing additional parameters or observational assumptions. The first examines the present deceleration parameter across the $(n,\Delta)$ plane and makes explicit a consequence of the fixed-$\Omega_{Q0}$ parametrization. The second considers the adiabatic response of the effective holographic component and clarifies the behavior of the corresponding background quantity near the phantom divide.
	
	\subsection{Present-Day Deceleration in the $(n,\Delta)$ Plane}
	\label{appsubsec:q0_parameter_plane}
	
	The two-dimensional behavior of the present deceleration parameter provides a useful structural check of the one-parameter results discussed in Sec.~\ref{subsubsec:q_transition}. At the present epoch, the relation between the Ricci-cutoff contribution and the deceleration parameter is
	\begin{equation}\label{eq:q0_appendix_relation}
		q_0=
		1-
		\left(
		\frac{\Omega_{DE0}}{\mathcal{C}_B}
		\right)^{\frac{2}{2-\Delta}}.
	\end{equation}
	For the two-dimensional map, the quantities $\Omega_{m0}$, $\Omega_{r0}$, $\Omega_{Q0}$, and $\mathcal{C}_B=\mathcal{C}_{B,\rm ref}$ are held fixed at their fiducial values, while $n$ and $\Delta$ are varied. The value $\mathcal{C}_{B,\rm ref}$ is calibrated once at the fiducial point using $q_{0,\rm ref}$ and is then kept fixed throughout the map; $q_0$ is therefore a derived quantity away from the fiducial point. Since
	\begin{equation}\label{eq:alpha_bar_fixed_OmegaQ_appendix}
		\bar{\alpha}(n)=\frac{\Omega_{Q0}}{2n-1},
	\end{equation}
	the power-law amplitude is adjusted for each $n$ so that the present geometric contribution remains fixed. The modified closure relation then fixes $\Omega_{DE0}$, and Eq.~(\ref{eq:q0_appendix_relation}) contains no explicit dependence on the power-law index $n$.
	
\begin{figure}[t]
		\centering
		\includegraphics[width=0.82\columnwidth]{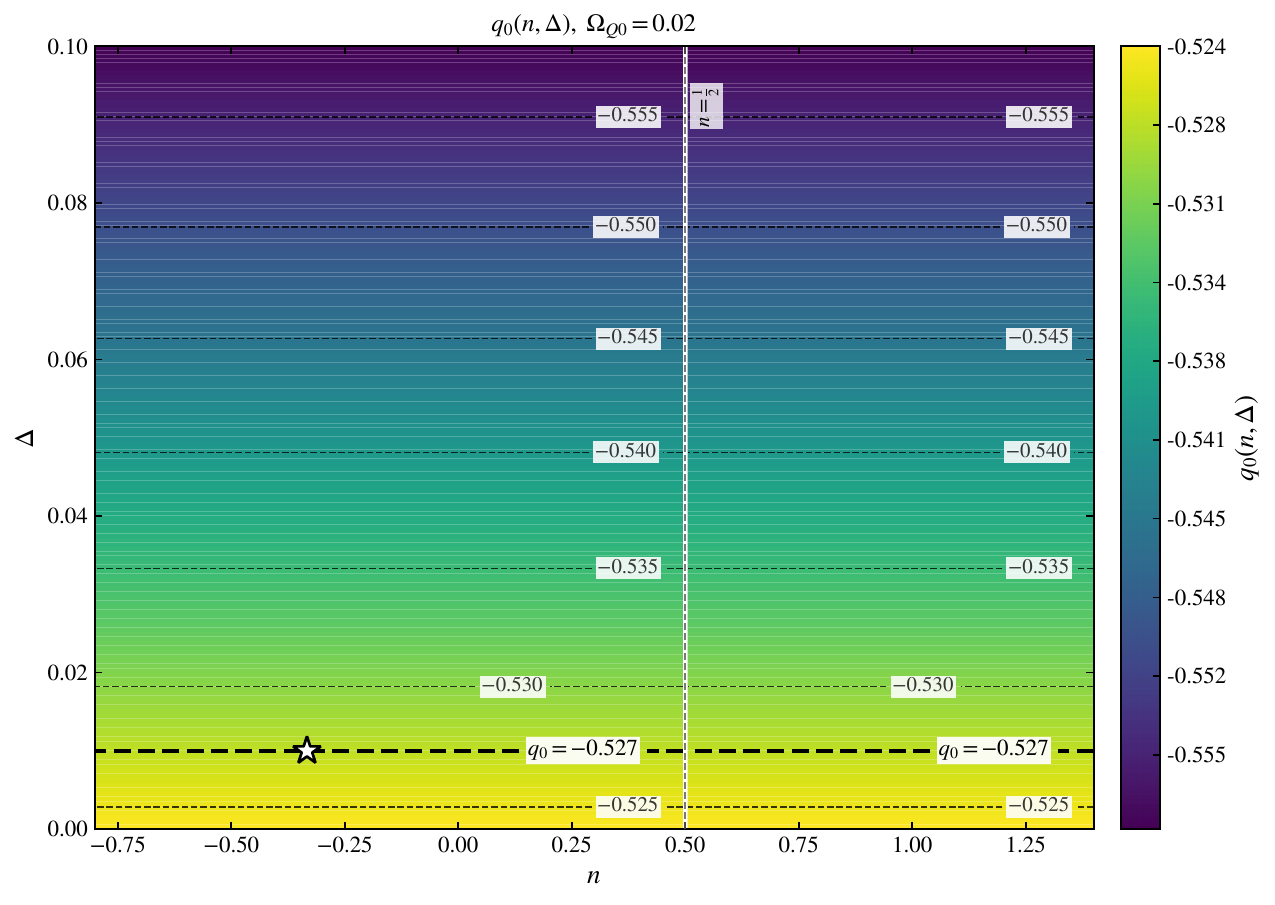}
		\caption{Present deceleration parameter $q_0$ in the $(n,\Delta)$ plane for the fixed-$\Omega_{Q0}$ parametrization adopted in the main analysis. The scan covers $-0.8\leq n\leq1.4$ and $0\leq\Delta\leq0.10$, with $\Omega_{m0}$, $\Omega_{r0}$, $\Omega_{Q0}$, and $\mathcal{C}_B=\mathcal{C}_{B,\rm ref}$ held fixed. The thick solid contour passes through the fiducial value $q_0\simeq-0.527$, the thinner dashed curves denote neighboring constant-$q_0$ contours, and the star marks the fiducial cosmology of Table~\ref{tab:fiducial}. The contours are horizontal to numerical precision because $q_0$ has no explicit $n$ dependence under this normalization. The vertical line at $n=1/2$ marks the value at which the mapping $\bar{\alpha}=\Omega_{Q0}/(2n-1)$ cannot represent a finite nonzero fixed $\Omega_{Q0}$; it is excluded from this particular scan parametrization rather than representing a singular background solution.}
		\label{fig:q0_parameter_plane_appendix}
	\end{figure}

	Figure~\ref{fig:q0_parameter_plane_appendix} displays this structure over the $(n,\Delta)$ plane. The star marks the fiducial cosmology, the thick solid contour passes through the fiducial value of $q_0$, and the thinner dashed curves trace neighboring constant-$q_0$ values. The constant-$q_0$ contours are horizontal to numerical precision, providing a direct graphical manifestation of the exact absence of an explicit $n$ dependence under the adopted fixed-$\Omega_{Q0}$ normalization.
	
	This behavior reproduces the one-dimensional $n$ scan summarized in Table~\ref{tab:representative_cosmological_results}. At fixed $\Delta=0.01$, all representative values of $n$ give the same present deceleration parameter to numerical precision,
	\begin{equation}\label{eq:q0_n_independence_appendix}
		q_0=-0.527316.
	\end{equation}
	The equality of these present-day values should not be interpreted as a general absence of dynamical sensitivity to $n$. As shown in the main analysis, changing the power-law index modifies the redshift evolution of the nonlinear geometric contribution and consequently changes $w_{DE}(z)$, the acceleration-transition redshift, the jerk parameter, and the higher-redshift expansion history.
	
	The dependence on $\Delta$ is qualitatively different. Along the fiducial-$n$ slice, with $\mathcal{C}_B=\mathcal{C}_{B,\rm ref}$ kept fixed after the fiducial calibration, increasing the Barrow deformation over the interval considered in Sec.~\ref{subsec:parameter_sensitivity} changes the present deceleration parameter from
	\begin{equation}\label{eq:q0_Delta_appendix_range}
		q_0=-0.524085
		\quad {\rm to}\quad
		q_0=-0.558265
	\end{equation}
	as $\Delta$ increases from $0$ to $0.10$. The color gradient in Fig.~\ref{fig:q0_parameter_plane_appendix} therefore reflects a genuine response of the present acceleration to the Barrow deformation within this fixed-$\mathcal{C}_B$ family.
	
	Figure~\ref{fig:q0_parameter_plane_appendix} thus explains an otherwise nontrivial feature of the one-dimensional sensitivity analysis. The constancy of $q_0$ along the $n$ direction follows from the chosen fixed-$\Omega_{Q0}$ parametrization and present-day normalization; it is not a general statement that the cosmology is independent of $n$. By contrast, $\Delta$ enters explicitly into the present Barrow--Ricci relation and changes the derived value of $q_0$ when the fiducial normalization $\mathcal{C}_{B,\rm ref}$ is held fixed.
	
	\subsection{Adiabatic Background Diagnostic}
	\label{appsubsec:adiabatic_sound_speed}
	
	A second supplementary probe is provided by the adiabatic quantity $c_a^2$ associated with the effective Barrow holographic component. As defined in Eq.~(\ref{eq:ca2_definition}),
	\begin{equation}\label{eq:ca2_appendix_definition}
		c_a^2\equiv\frac{\dot p_{DE}}{\dot\rho_{DE}},
	\end{equation}
	and, using the separate background conservation equation, its redshift representation is
	\begin{equation}\label{eq:ca2_appendix_redshift}
		c_a^2=
		w_{DE}
		+
		\frac{1+z}{3(1+w_{DE})}
		\frac{dw_{DE}}{dz},
	\end{equation}
	in agreement with Eq.~(\ref{eq:ca2_redshift}). We evaluate this quantity using the same numerical $w_{DE}(z)$ trajectories employed throughout the main background analysis. The derivative $dw_{DE}/dz$ entering this diagnostic is obtained from a cubic-spline reconstruction of the numerical equation-of-state trajectory; this spline-based prescription is used consistently throughout the calculation of $c_a^2$.
	
	It is useful to stress at the outset that $c_a^2$ is employed here strictly as a background diagnostic. In an effective cosmological fluid description it characterizes the ratio $\dot p_{DE}/\dot\rho_{DE}$ along the homogeneous solution, but it is not, in general, identical to the physical rest-frame propagation speed $c_s^2$ of scalar perturbations in the complete modified-gravity--holographic system. In general, $c_a^2\neq c_s^2$. The discussion below is therefore not intended as a substitute for a perturbative stability analysis.
	
\begin{figure*}[t]
		\centering
		\includegraphics[width=0.96\textwidth]{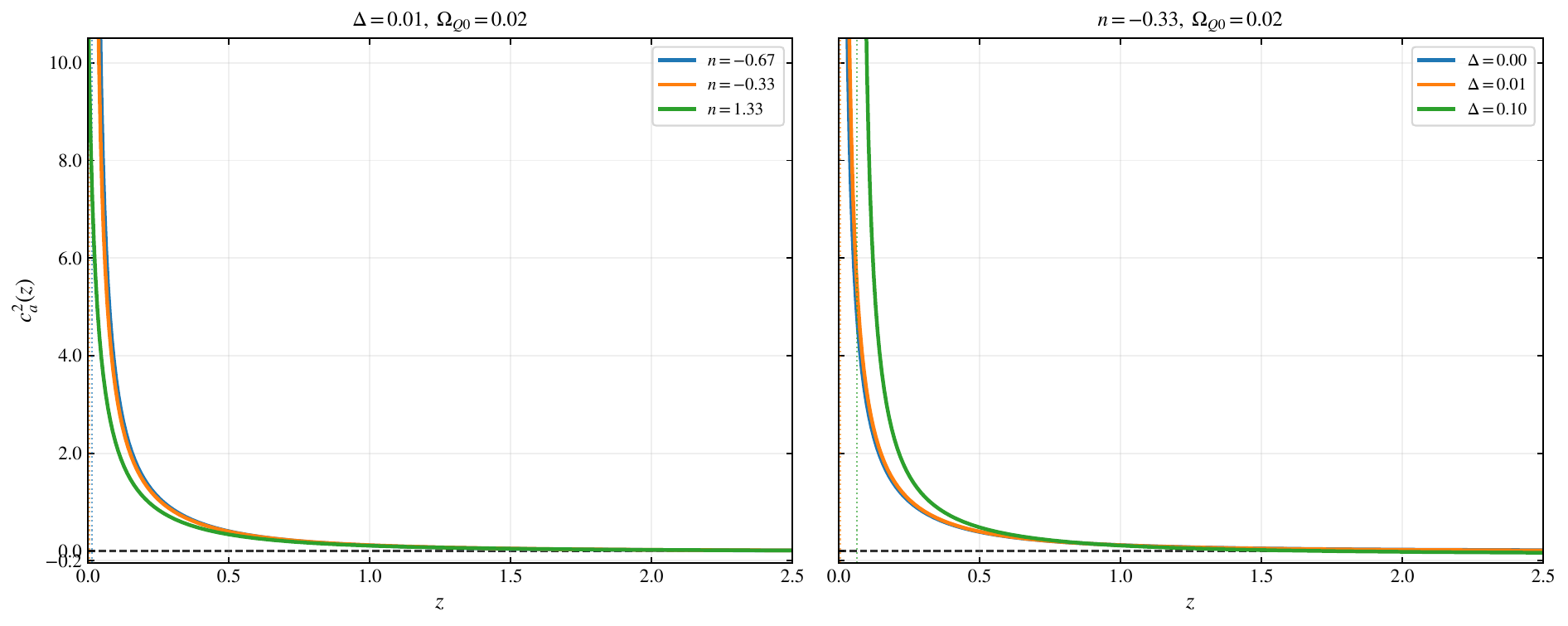}
		\caption{Evolution of the background adiabatic quantity $c_a^2(z)$ for representative parameter choices. The left panel shows $n=\{-2/3,-1/3,4/3\}$ for fixed $\Delta=0.01$, while the right panel displays $\Delta=\{0,0.01,0.10\}$ for fixed $n=-1/3$; all remaining background quantities are fixed according to Table~\ref{tab:fiducial}. The horizontal dashed line marks $c_a^2=0$, and the vertical dotted markers identify phantom-divide crossings satisfying $w_{DE}=-1$. Because the ratio defining $c_a^2$ becomes ill-conditioned close to the phantom divide, a narrow neighborhood satisfying $|1+w_{DE}|\leq2\times10^{-3}$ is masked in the plotted diagnostic. The displayed vertical range, $-0.25\leq c_a^2\leq10.5$, is chosen to emphasize the finite branches away from this masked neighborhood. This masking affects only the displayed background diagnostic and does not modify the numerical integration of the background solution.}
		\label{fig:ca2_appendix}
	\end{figure*}

	Figure~\ref{fig:ca2_appendix} shows the redshift evolution of $c_a^2$ for representative values of $n$ and $\Delta$. The left panel displays $n=\{-2/3,-1/3,4/3\}$ at fixed $\Delta=0.01$, while the right panel shows $\Delta=\{0,0.01,0.10\}$ at fixed $n=-1/3$. The remaining background quantities are fixed to the fiducial values. The horizontal dashed line denotes $c_a^2=0$, while the vertical dotted markers indicate the locations of the corresponding phantom-divide crossings, $w_{DE}=-1$.
	
	The most prominent feature of Fig.~\ref{fig:ca2_appendix} occurs near the phantom divide. The redshift expression for $c_a^2$ contains the factor $(1+w_{DE})^{-1}$ and can therefore become formally undefined or divergent as $w_{DE}\rightarrow-1$. At an exact crossing, the separate conservation equation gives $\dot\rho_{DE}=0$, so the defining ratio in Eq.~(\ref{eq:ca2_appendix_definition}) can likewise become ill-defined. This behavior belongs to the derived adiabatic ratio itself and does not imply a singularity of the homogeneous cosmological solution. The numerical evolution of $E(z)$, $w_{DE}(z)$, and the remaining background quantities remains smooth through the crossing.
	
	For the fiducial cosmology, the crossing occurs very close to the present epoch,
	\begin{equation}\label{eq:fiducial_phantom_crossing_appendix}
		z_{w=-1}\simeq0.0066,
	\end{equation}
	whereas for the representative case $\Delta=0.10$ it is shifted to
	\begin{equation}\label{eq:Delta_high_phantom_crossing_appendix}
		z_{w=-1}\simeq0.065.
	\end{equation}
	These values are consistent with the equation-of-state evolution discussed in Sec.~\ref{subsubsec:wde_evolution}.
	
	Away from the immediate neighborhood of the phantom divide, the adiabatic quantity rapidly returns to finite values. For the fiducial trajectory, representative values are
	\begin{equation}\label{eq:ca2_fiducial_selected_values}
		c_a^2(0.5)\simeq0.40,
		\qquad
		c_a^2(2)\simeq0.022.
	\end{equation}
	The trajectories corresponding to different values of $n$ remain relatively close at intermediate and high redshift, consistently with the mild $n$ dependence found for several background observables in the main analysis.
	
	The dependence on the Barrow deformation is more pronounced. For $\Delta=0$ and $\Delta=0.01$, the diagnostic remains slightly positive near $z=2$, with
	\begin{equation}\label{eq:ca2_small_Delta_z2}
		c_a^2(2)\simeq0.026
		\quad {\rm and}\quad
		c_a^2(2)\simeq0.022,
	\end{equation}
	respectively. For $\Delta=0.10$, the same background diagnostic becomes slightly negative,
	\begin{equation}\label{eq:ca2_large_Delta_z2}
		c_a^2(2)\simeq-0.019.
	\end{equation}
	This sign change characterizes the background adiabatic response of the effective holographic sector and should not, by itself, be identified with a fundamental gradient instability.
	
	In particular, neither a negative value nor the formal divergence of $c_a^2$ near a phantom-divide crossing determines the physical scalar propagation speed or the perturbative stability of the complete system. A definitive assessment would require the linear perturbation equations together with the corresponding kinetic and gradient terms, from which the physical sound speed and no-ghost and gradient-stability conditions could be derived. Those perturbative quantities are beyond the scope of the present background analysis.
	
	The two supplementary diagnostics therefore clarify complementary structural aspects of the model. Figure~\ref{fig:q0_parameter_plane_appendix} makes explicit the normalization-induced absence of an $n$ dependence in the present deceleration parameter under the fixed-$\Omega_{Q0}$ prescription, while Fig.~\ref{fig:ca2_appendix} illustrates the background adiabatic response and its formal sensitivity to the phantom divide. The latter is evaluated consistently with the spline-based prescription described above and is not used as an independent validation of the background solver or as a perturbative stability criterion.
	

	\bibliography{Ref_tavakoli}
	
\end{document}